\documentclass[twocolumn]{aastex63}

\usepackage{hyperref} 

\usepackage{amsmath,amsopn,amsxtra} 
\usepackage{comment} 
\usepackage{float} 
\usepackage{natbib} 
\usepackage{threeparttable}
\usepackage{xcolor} 
\usepackage{soul}

\newcommand{\kms}{\ensuremath{\textrm{km\thinspace s}^{-1}}}
\newcommand{\K}{\ensuremath{ \, \mathrm{K}}}
\newcommand{\pc}{\ensuremath{ \, \mathrm{pc}}}
\newcommand{\kpc}{\ensuremath{\, \mathrm{kpc}}}

\newcommand{\ang}{\ensuremath{ \, \mathrm{\AA}}}
\newcommand{\R}{\ensuremath{ \, \mathrm{R}}}

\newcommand{\cmsq}{\ensuremath{\textrm{cm}^{-2}}}
\newcommand{\hi}{H\textsc{~i}}
\newcommand{\SLB}{UVQSJ204402.02-075810.0}
\newcommand{\SLA}{UVQSJ203335.89-032038.5}
\newcommand{\SLE}{RXJ2139.7+0246}
\newcommand{\SLC}{RXJ2043.1+0324}
\newcommand{\SLD}{PG2112+059}

\newcommand{\cloudy}{\textsc{~Cloudy}}
\defcitealias{2016ApJ...816L..11F}{F16}

\definecolor{alexcolor}{rgb}{0.188, 0.858, 0.478}

\definecolor{franciecolor}{rgb}{0.72, 0.53, 0.04}

\newcommand{\shSLA}{$-0.46^{+0.38}_{-0.37}$}
\newcommand{\sisSLA}{$-0.72^{+0.24}_{-0.26}$}
\newcommand{\sihSLA}{$-1.19^{+0.34}_{-0.35}$}

\newcommand{\sioSLA}{$-0.05$}

\newcommand{\shSLB}{$+0.28^{+0.25}_{-0.26}$}

\shorttitle{Smith Cloud}
\shortauthors{Vazquez et al.}

\begin{document}

\title{Dust in the Smith Cloud? A UV Investigation into the Smith Cloud's Gas-Phase Abundance Patterns 
}

\author[0000-0003-2308-8351]{Johanna T. V{\'a}zquez}
\affiliation{Department of Physics \& Astronomy, Texas Christian University, Fort Worth, TX 76129, USA}

\author[0000-0001-5817-0932]{Kathleen A. Barger}
\affiliation{Department of Physics \& Astronomy, Texas Christian University, Fort Worth, TX 76129, USA}

\author[0000-0003-4237-3553]{Frances H. Cashman}
\affiliation{Presbyterian College, Clinton, SC 29325, USA}
\author[0000-0003-0724-4115]{Andrew J. Fox}
\affiliation{AURA for ESA, Space Telescope Science Institute, Baltimore, MD 21218}
\affiliation{Department of Physics \& Astronomy, Johns Hopkins University, 3400 N. Charles Street, Baltimore, MD 21218, USA}

\author[0000-0002-0507-7096]{Bart P. Wakker}
\affiliation{Supported by NASA/NSF, affiliated with Department of Astronomy, University of Wisconsin-Madison, Madison, WI 53706, USA}

\author[0000-0002-6050-2008]{Felix J.\ Lockman}
\affiliation{Associated Universities Inc., Washington, DC 22180, USA}
\affiliation{Green Bank Observatory, Green Bank, WV 24944, USA}

\author[0000-0001-7301-5666]{Alex S. Hill}
\affiliation{Department of Computer Science, Mathematics \& Physics, University of British Columbia-Okanagan, Kelowna, BC V1V 1V8, Canada}
\affiliation{Dominion Radio Astrophysical Observatory, Herzberg Astronomy \& Astrophysics Research Centre, National Research Council Canada, Penticton, BC V0H 1K0, Canada}

\author[0000-0003-0536-3081]{Suraj Poudel}
\affiliation{Department of Physics \& Astronomy, Texas Christian University, Fort Worth, TX 76129, USA}

\author[0009-0005-3076-1104]{April L. Horton}
\affiliation{Department of Physics \& Astronomy, Texas Christian University, Fort Worth, TX 76129, USA}

\author{Jaq Hern{\'a}ndez}
\affiliation{Space Telescope Science Institute, Baltimore, MD 21218, USA}
\affiliation{Department of Physics \& Astronomy, Texas Christian University, Fort Worth, TX 76129, USA}

\author{Matthew Nuss}
\affiliation{Department of Engineering, University of Washington, Seattle, WA 98195, USA}
\affiliation{Department of Physics \& Astronomy, Texas Christian University, Fort Worth, TX 76129, USA}

\author{Alice Blake}
\affiliation{University of Vermont, Burlington, VT 05405, USA}
\affiliation{Department of Physics \& Astronomy, Texas Christian University, Fort Worth, TX 76129, USA}

\author[0000-0002-0646-1540]{Lauren Corlies}
\affiliation{Dominion Radio Astrophysical Observatory, Herzberg Astronomy and Astrophysics Research Centre, National Research Council Canada Box 248, Penticton, BC V2A 6K3, Canada}

\author[0000-0003-1455-8788]{Molly Peeples}
\affiliation{Space Telescope Science Institute, Baltimore, MD 21218, USA}
\affiliation{Department of Physics \& Astronomy, Johns Hopkins University, 3400 N. Charles Street, Baltimore, MD 21218, USA}

\begin{abstract}
The Smith Cloud is a high-velocity cloud (HVC) on its final approach to the Milky Way that shows evidence of interaction with the Galaxy's disk. We investigate the metallicity and gas-phase chemical depletion patterns in this HVC using UV absorption-line observations toward two background QSOs taken with the Hubble Space Telescope (\textit{HST})/Cosmic Origin Spectrograph (COS) and H\textsc{~i} 21-cm emission-line observations taken with Green Bank Telescope (GBT). We find evidence of silicon gas-phase depletion with [Si/S]~=~\sisSLA\, and $\text{[Si/O]}_{3\sigma}~ \lesssim$~\sioSLA, implying the presence of dust within the Smith Cloud. Because dust is galactic in origin, this HVC could trace the return leg of a Galactic fountain or a dwarf galaxy that passed through the Galactic plane.
\end{abstract}


\section{Introduction}\label{section:intro}

High-velocity clouds (HVCs) are gas clouds in the circumgalactic medium of galaxies that move at velocities inconsistent with the rotation of their host galaxy. For the Milky Way's HVC population, this is typically defined as $| v_{\rm{LSR}}|>90\,\kms$ \citep{1997ARA&A..35..217W}. These HVCs have a variety of origins, from galactic fountains, satellite galaxy interactions, intergalactic medium, and condensation of galactic halo gas (see \citealt{1997ARA&A..35..217W}, \citealt{2012ARA&A..50..491P}, \citealt{2017ASSL..430...15R}, and references therein). Their origins are generally assessed by their positions, motions, morphologies, and compositions. 

Both metallicity and the presence or lack of dust together provide significant constraints on the origin and history of HVCs. Clouds with high metallicity are generally associated with either the Milky Way ($Z\gtrsim0.5\,Z_{\odot}$; e.g., \citealt{2016ApJ...816L..11F, 2023ApJ...944...65C}) or the Magellanic Clouds (MCs) ($Z\gtrsim0.3\,Z_{\odot}$; e.g., \citealt{2013ApJ...772..111R}) when accounting for their positions and motions. Although there are low-metallicity HVCs that are attributed to a MC origin (e.g., \citealt{2008ApJ...678..219L,2013ApJ...772..110F, 2018ApJ...854..142F}), in general, the low metallicity HVC clouds that surround the MW have more ambiguous origins. The presence of dust in an HVC provides an additional constraint on whether or not it could have a galactic origin. This is because interstellar dust primarily forms in the surfaces of asymptotic giant branch (AGB) stars and during supernovae (\citealt{2006ApJ...638..262F}, \citealt{2011ApJ...727...63D}). Stellar feedback events cause the interstellar medium (ISM) to be polluted with dust, which can then be launched into the circumgalactic medium (CGM) via galactic feedback. Studies in the far-infrared (FIR) have attempted to characterize dust in the ISM, CGM, and intergalactic medium (IGM; e.g. \citealt{1986A&A...170...84W}, \citealt{2016A&A...586A.121L}, \citealt{2024MNRAS.tmp..312H}.  However, for HVCs, there are significant limits to the sensitivities of these measurements, which we discuss in Section \ref{section:discussion}.  Measurement of gas phase $\alpha$ element depletion patterns has been an important alternative method of inferring the presence of dust (e.g. \citealt{2001ApJ...559..318R}), as it is sensitive to small changes in column density. 

One particular HVC that has been studied extensively, yet has an origin shrouded in mystery, is the Smith Cloud (SC). Discovered in 1963 \citep{1963BAN....17..203S},
this HVC has spawned great interest due to its trajectory. \cite{2008ApJ...679L..21L} found that the SC is on a collision course with the Milky Way's Galactic plane based on this cloud's cometary morphology, velocity distribution, and close position below the Galactic disk. The tip of the SC lies at a distance from the Sun of $d_\odot=12.4\pm1.3\,\kpc$ and $z=-2.9\pm0.3\,\kpc$ at $(l,\,b)=(39\arcdeg,\,-13\arcdeg)$ from the Galactic plane \citep{2003ApJ...597..948P, 2008ApJ...679L..21L, 2008ApJ...672..298W}. Furthermore, \cite{2023ApJ...943...55L} found that the SC could already be in the early stages of collision with the Milky Way's Galactic plane. 

\citet{2009ApJ...703.1832H}, using optical emission-line observations, characterized the chemical composition of the SC's main body to be $-1.31 \lesssim \text{[N/H]} \lesssim -0.40$, based on the observed line ratio of $[N\textsc{~ii}]~\lambda6583/H\alpha=0.32\pm0.05$, corresponding to a nitrogen abundance of [N/H]~$=~-0.82\pm0.06$(also see \citealt{1998MNRAS.299..611B}). 
\cite{2016ApJ...816L..11F} (F16) used UV absorption, \hi~21-cm emission, and photoionization modeling to estimate that this HVC has an average sulfur abundance of $[{\rm S/H}]=-0.28\pm0.14$ in the trailing gas.  This discrepancy between lower and higher metallicity measurements may be resolved if this HVC is experiencing substantial ram-pressure stripping that has promoted the mixing of the outer layers with a high-metallicity Galactic halo, increasing the metallicity in the trailing gas \citep{2017ApJ...837...82H}. However, F16 argued that their metallicity distribution was consistent with a decreasing metallicity gradient from the main body of the SC to the tail.  However, they did not have sufficient data to make a confident determination of a metallicity gradient. More measurements of chemical abundances across the SC, particularly in and near its main body, are necessary to resolve this discrepancy and to understand the metallicity distribution within this intriguing HVC. 

This study provides two additional UV-based sulfur abundance measurements within the SC and the first measurements of gas-phase silicon depletion in the SC. For this work, we use \textit{HST}/COS/G130M UV absorption-line observations toward five bright background QSOs and GBT \hi\ 21-cm emission-line observations, which we describe in Section~\ref{section:observations}.  In Section~\ref{section:specanalysis}, we outline our methodology for determining the spectral parameters, such as kinematic centroids, kinematic widths, and column densities. In Section~\ref{section:cloudy}, we describe our procedure constraining chemical abundance patterns of this HVC using photoionization simulations anchored on the observations. In Section~\ref{section:abundances} we discuss our analysis and results for our abundance patterns. We then discuss all of our results and their implications to the history and the future of the SC in Section~\ref{section:discussion} and summarize our findings in Section~\ref{section:summary}. 

\section{Observations \& Reduction}\label{section:observations}
We utilize UV absorption-line observations taken with HST's Cosmic Origins Spectrograph (COS). This includes archival \textit{HST}/COS observations from program \textit{HST}-13840 (PI: Fox, \citealt{2016ApJ...816L..11F}) along three sightlines and new \textit{HST}/COS observations come from program \textit{HST}-15161 (PI: Barger) along two sightlines. We additionally use \hi~21-cm emission-line observations taken with the Green Bank Telescope (GBT). We use existing \hi\, maps from \citep{2008ApJ...679L..21L} and deep-pointed observations for all five sightlines from a new GBT survey by \cite{LockmanInPrep}. Here we also describe the reduction techniques we employed for both the UV absorption and H\textsc{~i} emission data. 

\subsection{UV Absorption}\label{subsection:uv_reduction}

We acquired new \textit{HST}/COS observations using the G130M grating centered at $\lambda=1291\,\ang$ (PI: Barger) along background QSOs \SLA\, (referred to henceforth as Sightline~D; see Table \ref{table:targets}) and \SLB\, (referred to henceforth as Sightline~E; see Table \ref{table:targets}). The COS/G130M/1291 grating has a spectral resolution ranging from $R = 12{,}500$ at 1150\ang\; to $R=16{,}500$ at 1450\ang\; \citep{2022cosd.book..5.1S}, corresponding to a FWHM which ranges from $17.7\,\kms$ at the high wavelength end to $24.8\,\kms$ at the low wavelength end. The $1300$\ang\, region of the spectrum is contaminated by geocoronal airglow emission. Therefore, when necessary, we measured the apparent column densities of O\textsc{~i}$\lambda\, 1302$ and {\rm Si}\textsc{~ii}~$\lambda\,1304$ using night-only observations. The night-only data correspond to periods when the telescope was in Earth's shadow and is extracted by selecting the time intervals when the Sun angle as seen by the telescope is less than $20^{\circ}$. To do this, we ran the CalCOS data reduction pipeline twice, once selecting only the night-only data and once selecting all data (day plus night). 
\begin{center}
\begin{deluxetable*}{c c c c c c c c}\label{table:targets}
\tablecaption{ A table of Galactic coordinates, LSR velocities, Doppler velocity widths ($b$), and logarithmic \hi\, column densities ($\log N_{\rm{\hi}}$).}
\tablecolumns{7}
\tablewidth{0pt}
\tablehead{
\colhead{ID} \vspace{-0.2cm} & QSO & Source & \colhead{$\ell_{\rm{gal}}$} & \colhead{$b_{\rm{gal}}$}&  \colhead{$v_{\rm{LSR}}$} & $b$&\colhead{$\log N_{\rm{H\textsc{~i}}}$} \\ \colhead{}  & \colhead{} & \colhead{}  & \colhead{$(^\circ)$} &\colhead{$(^\circ)$} &\colhead{(\kms)}&\colhead{(\kms)} & \colhead{(\cmsq)}  \\
}
\startdata
A & \SLC & F16 & $+49.72$ & $-22.88$ & $+81.8$ & $14.8$ & $18.67$\\
B\tablenotemark{a} & \SLD & F16 & $+57.04$ & $-28.01$ & $+54.3$ & $12.1$ & $18.38$\\
C & \SLE & F16 & $+58.09$ & $-35.01$ & $+58.1$ & $15.1$ & $19.32$\\
 D &\SLA & This Study & $+42.04$  &  $-24.18$ &   $+46.6,+114.7$\tablenotemark{c} & $5.0, 24.6$\tablenotemark{c} & $19.22, 18.56$\tablenotemark{c}  \\   
E\tablenotemark{b} &\SLB & This Study & $+38.79$& $-28.62$ &  $+63.0$ & $17.8$ & $19.37$   \\ 
\enddata
\tablenotetext{a}{We re-calculate the \hi\ 21-cm fitting parameters for this sightline~based on new GBT observations.}
\tablenotetext{b}{This sightline requires a three-component fit (shown in Table \ref{table:profile_fits}), which we detail in Section \ref{subsection:HI_Analysis}. We report the LSR velocity as the column density-weighted average. For compactness, report the Doppler $b$ values added in quadrature, the co-added column density, and the column-density-weighted mean of centroid velocity.}
\tablenotetext{c}{This component had spectral abnormalities, which we discuss in Section \ref{subsection:hi_reduction}}
\label{table: sightlines}
\end{deluxetable*}
\end{center}
\subsection{\hi~21-cm Emission}\label{subsection:hi_reduction}

\begin{center}
    
\begin{figure*}
    \centering
    \includegraphics[width=14cm]{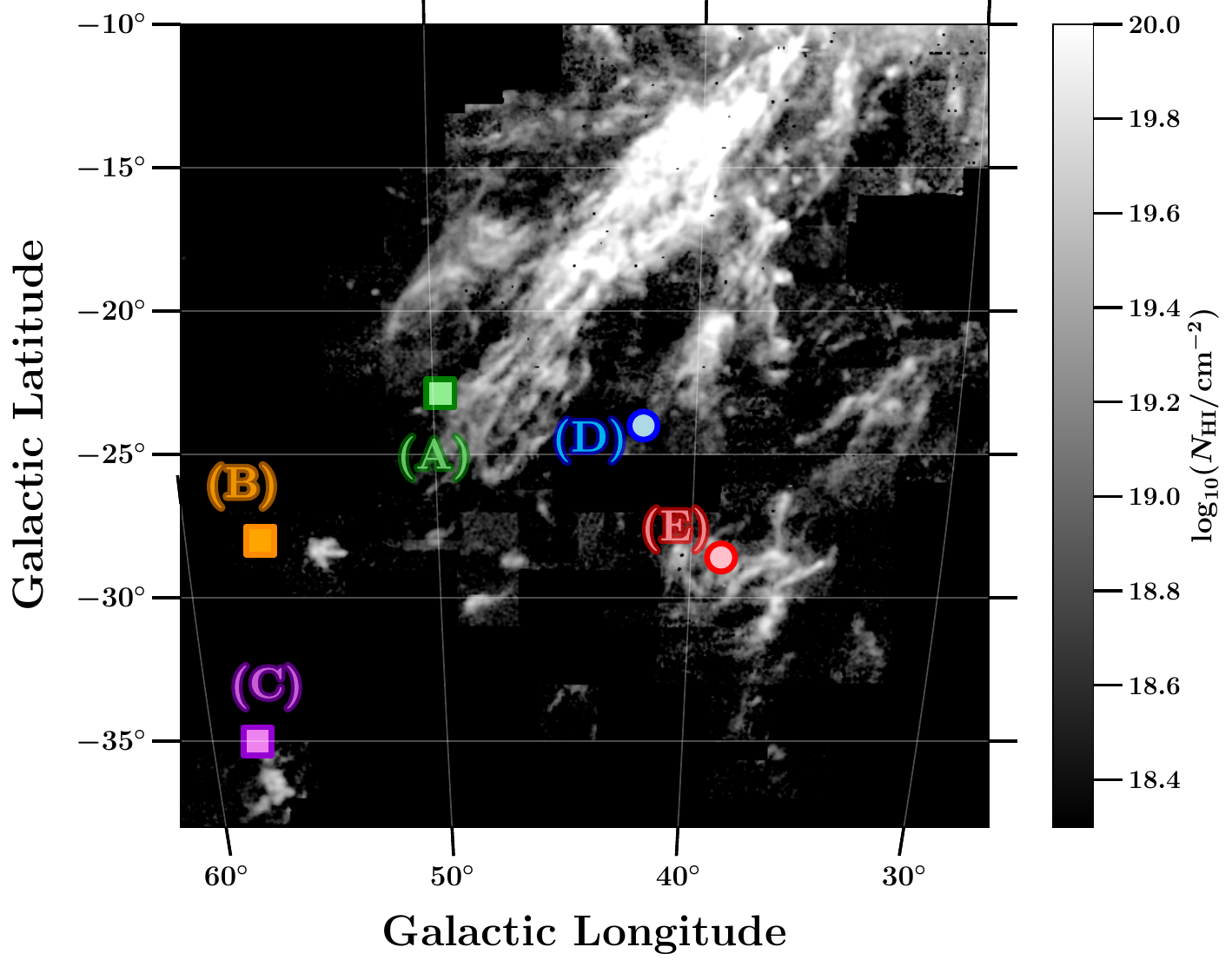}
    \caption[SC Sightline Map]{The distribution of our UV background targets overlayed with a kinematically-integrated H\textsc{~i} emission map with $v_{\rm{LSR}} \geq +66\,\kms$  \citep{2008ApJ...679L..21L}. The circle markers refer to targets probing \hi\ gas along cloud fragments adjacent to the SC's main body. The square markers refer to targets probing \hi\ gas in the trailing wake of this HVC. We label each target in this figure with its simplified name (see Table \ref{table: sightlines}).} 
    \label{fig:SC_map}
\end{figure*}
\end{center}

We obtained new pointed GBT \hi~21-cm emission-line observations along our five sightlines (PI: Wakker; PID: GBT23A-344). We took four 600~second exposures in frequency switched mode, with a 2\,MHz (400\,\kms) offset from the central frequency.  We applied a stray-radiation correction to the mirrored and subtracted spectra following the method outlined in the \citet{2011A&A...536A..81B} study. We note that because emission at velocities above $v_{\rm{LSR}}\approx +50\,\kms$ is generally weak, and because the shape of the GBT dish suppresses sidelobes, this correction is basically negligible for the IVC/HVC emission.  

For sightline~A, B, C, and~E, we fit a high-order (20th) polynomial to the resulting baseline, separately for the XX and YY polarizations, in order to remove the residual offsets, which are on the order of 0.05~K. We also filtered out a few 0.005~K interference spikes that occurred in three fixed channels. After fitting the baseline for the XX and YY polarizations for each of the four exposures, we averaged the resulting eight spectra together. The original data had a velocity resolution of $\Delta v_{\rm{res}}=0.3\,\kms$, which we smoothed to $\Delta v_{\rm{res}}=1.2\,\kms$. The RMS of the baseline of the  final spectra is roughly 0.0040 K, corresponding to a $3\sigma$ detection limit of about $N_{\rm{\hi}}~=~4\times$10$^{17}$\,\cmsq. 

For Sightline~D, when we inspected XX and YY polarizations, we found that their continua differed. We were able to fit the continuum of the XX polarization data with linear regression. However, for the YY polarization dataset and the combined average spectrum of both polarizations---the latter of which we use to calculate \hi\ column densities---we had to use a 5-order polynomial to fit its continuum. This disagreement between the polarization datasets for Sightline~D translates to a large uncertainty in the baseline continuum, especially in the velocity range $+60 \lesssim v_{\rm{LSR}} \lesssim +80\,\kms$. This impacted our \hi\ component fits, which we discuss in Section~\ref{SLD_HI}.  We report our findings in the \hi\, spectral analysis with a geometric average of the column densities. We supplement our pointed observations with H\textsc{~i}\,21-cm datacubes from \cite{2008ApJ...679L..21L}.

\section{Spectral Analysis}\label{section:specanalysis}

In this section, we discuss our methodology for analyzing UV absorption and \hi~21-cm emission spectra. First, in Section  \ref{subsection:HI_Analysis}, we outline our approach to determining \hi\ column densities and kinematic parameters, along with the resulting measurements. Next, in Section \ref{subsection: UV_Absorp}, we describe our methodology for assessing column densities and kinematic properties of our lowly ioized UV lines, highlighting comparisons and discrepancies with the \hi\ spectra. Finally, in the remaining subsections, we explore the peculiarities encountered while fitting these spectra."

\subsection{H\textsc{~i}\,21-cm Emission}\label{subsection:HI_Analysis} 

To determine the H\textsc{~i} column density, we utilize the following relation with brightness temperature ($T_B$) \citep{1990ARA&A..28..215D}:
\begin{equation}
    N_{\text{H\textsc{~i}}} = 1.823 \times 10^{18} \;\frac{\text{atoms}}{\text{cm}^2} \int_{v_{\text{min}}}^{v_{\text{max}}} \frac{T_B(v)dv}{\rm K},
    \label{eqn:hi}
\end{equation}
which assumes that the cloud is transparent to 21-cm light such that all the emission from the cloud is able to escape.
We characterized the kinematic properties of the \hi\, gas in the SC by Gaussian fitting the emission using the \texttt{gaussdecomp} Python package \citep{gaussdecomp}. This library provides a best-fit multiple-Gaussian fit to an \hi~21-cm spectrum while varying the number of Gaussians as a free parameter. For sightlines A and C, we have new \hi\, measurements from GBT23A-344, and we find column densities consistent with those measured by F16. We will discuss the \hi\, analysis for sightlines B, D, and E in more detail below.

\subsubsection{Sightline B}
Although F16 analyzed \hi\ data from the Leiden-Argentine Bonn (LAB; \citealt{2005A&A...440..775K}) Survey for sightline~B, GBT observations have a factor of 15 smaller beamsize than LAB. The GBT observations also have an order of magnitude greater signal-to-noise ratio than LAB observations, improving our ability to detect faint \hi\ 21-cm emission.  F16 previously measured an \hi\, column density of $\log{(N_{\rm{\hi}}/\cmsq)} = 18.72\pm 0.06$ along this sightline. Our deep-pointed GBT observations yield a component at $v_{\rm{LSR}} = +54.3\pm1.0\,\kms$ with a Doppler width of $b=12.1\pm1.3\,\kms$ and a column density of $\log\left(N_{\rm{\hi}}/\cmsq\right)= 18.37\pm0.09$. Due to this large discrepancy in \hi\, column density measurements, we re-calculated the metallicity and ionization parameter along this sightline using the new \hi\, column density measurements. We use the S\textsc{~ii} column densities from F16 and run photoionization models at the new \hi\ column density (see Section~\ref{subsection:pg2112_opt}).
\subsubsection{Sightline D}\label{SLD_HI}
Toward Sightline~D, we find two  kinematic   components  that may be associated with the SC at $v_{\rm{LSR}} = +46\,\kms$ and $v_{\rm{LSR}} = +115\,\kms$ (see Figure~\ref{fig:SC_vel_chan_maps}). However, the \hi\ at this lower velocity does not share the SC's characteristic cometary morphology, and the kinematic width of the gas along this sightline  is much more narrow at $b_{\rm{HI}}=5\,\kms$ than the other absorption features along all of the explored sightlines (see Table~\ref{table:profile_fits}). Therefore, this component may instead be associated with an unresolved IVC or the Sagittarius Arm of the Milky Way (SAMW). The higher velocity component lies along a cloud fragment that has an elongated morphology that aligns with the main body of the SC (see Figure~\ref{fig:SC_vel_chan_maps}) and is therefore more likely to be associated with this HVC.

Ascertaining the associated column density of the \hi\ emission for this high velocity component relies on an adequate characterization of the spectral continuum.  In the XX polarization dataset for our pointed GBT observations, we  find elevated signal between $+55 \lesssim v_{\rm{LSR}\lesssim}+80\,\kms$ that we do not observe along neighboring sightlines in the \citet{2008ApJ...679L..21L} mapped GBT observations. We also find a significant difference in measured \hi\, column densities depending on the polarization for which we make the measurement. For the XX polarization dataset, we find a column density of $\log(N_{\rm{\hi}}/\cmsq) = 18.81\pm0.01$ and $\log(N_{\rm{\hi}}/\cmsq) = 18.32\pm0.05$ using the YY polarization dataset.
This discrepancy adds uncertainty to our continuum fit, which we reflect in the elevated uncertainty in the fit parameters. We report the geometric mean of the fit of the XX spectrum and the fit of the YY spectrum, which yields a  column density of $\log(N_{\rm{\hi}}/\rm{cm^{-2}}) = 18.56\pm 0.25$ (including deviation from individual polarizations) and a Doppler parameter kinematic width of $b = 24.6\pm2.3\,\kms$ using the continuum fitting techniques that we discuss in Section~\ref{subsection:hi_reduction}.

\subsubsection{Sightline E} \label{subsubsection: SLE_HI_analysis}
\begin{figure*}
    \centering 
    \includegraphics[width=14cm]{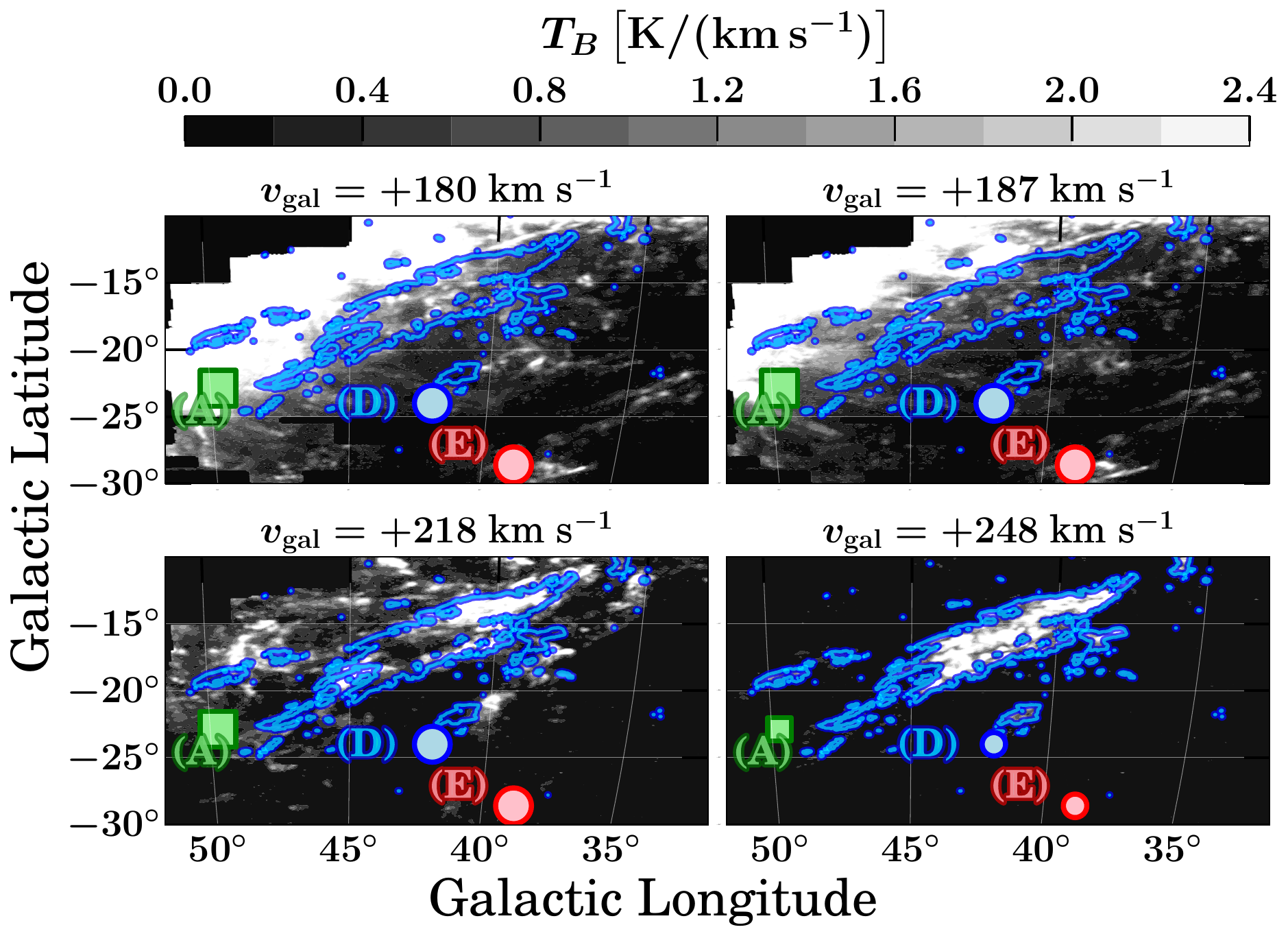}
    \caption{All panels are \hi~21-cm velocity channel maps of the SC in galactocentric standard of rest (GSR) velocities ($v_{\rm{GSR}} = v_{\rm{LSR}} + V_{\rm{tan}}\sin \ell\cos b)$. The color axis measures the brightness temperature bins at different central velocities with bin velocity of $\Delta v = 1.2\,\kms$. We distinguish the GSR  velocities of the \hi\, channel maps at the top of each respective panel. We additionally highlight the coordinates of sightlines A, D, and E in each panel. The blue contour traces gas at $v_{\rm{GSR}} = +248.4\,\kms$  with a brightness temperature greater than $0.3\,\rm{K}/(\kms)$.}
    \label{fig:SC_vel_chan_maps}
\end{figure*}

At the velocity channel $v_{\rm{LSR}}=+63.7\,\kms$, Galactic longitude $\ell\approx+35^\circ$, and Galactic latitude $-25^\circ ~\lesssim b \lesssim -20^\circ $, there is an \hi\ filament with an angle of attack and cometary structure consistent with that of the SC (see Figure~\ref{fig:SC_vel_chan_maps}). However, material at these latitudes at velocities of $v_{\rm{LSR}}\approx +65\,\kms$ is \textit{also} consistent with the SAMW \citep{Vall_e_2017}. We must then emphasize that the origins of the cloud are ambiguous. It is also possible that we have heterogeneous origins for the various \hi\ components, further confusing any results that we could infer for the SC. Although we report our analysis of the gas along this sightline, we are cautious about any inferences we could make regarding the physical conditions of the SC.

The asymmetric \hi\ emission at $v_{\rm{LSR}} = +65\,\kms$ is best fit by three separate Gaussian components with an average velocity of $v_{\rm{LSR}} = +63.0\pm 2.4\,\kms$ (see Table~\ref{table:profile_fits}). However, the kinematic resolution of \textit{HST}/COS is insufficient to resolve this multiple component structure. For the sake of consistency, we co-add these \hi\ components to obtain an column density of $\log\left(N_{\rm{\hi}}/\cmsq\right) = 19.37\pm0.02$.


\subsection{UV Absorption}\label{subsection: UV_Absorp}
To characterize the properties of the SC, we use \textit{HST}/COS UV absorption spectra taken with the G130M grating centered at $1291\,\ang$. To estimate column densities, we utilize two methods. For the saturated and low S/N cases, we use the apparent optical depth (AOD; \citealt{1991ApJ...379..245S}) method to determine  column density limits. We find these limits by integrating the AOD over some velocity: 
\begin{equation}\label{eqn:aod_colden}
    \frac{N_a}{\text{\cmsq}} = \frac{3.768\times 10^{14}}{f_{ij}\lambda_{ij}/\rm{\ang}}\int_{v_{\rm{min}}}^{v_{\rm{max}}}\ln\left(\frac{F_{\rm{cont}}(v)}{F_{\rm{obs}}(v)}\right) dv,
\end{equation}
where $N_a$, $f_{ij}$, and $\lambda_{ij}$ respectively represent the apparent column density, oscillator strength, and central wavelength of an ionic transition $i\xrightarrow{}j$. For lines with undetected absorption, we report $3\sigma$ upper limits on their AOD column densities. For saturated lines, we report the AOD column densities as lower bounds.

For the case of unsaturated absorption in at least one line transition, we incorporate Voigt line-profile fitting.  We use the \texttt{VoigtFit} Python package \citep{krogager2018} to fit the absorption features and to calculate column densities, centroid velocities, and line widths. This package utilizes atomic transition data (e.g., \citealt{2003ApJS..149..205M}, \citealt{2004PhyS...69..196J}, \citealt{2017ApJS..230....8C}) for oscillator strengths, central wavelengths, and Einstein coefficients.\, \texttt{VoigtFit} also allows for co-fitting of multiple ions and multiple transitions of the same ion. This is especially useful when multiple transitions of the same ion are available, as co-fitting can help to reduce the uncertainties that would otherwise occur when some (not all) of the absorption features lie on the flat portion of the curve of growth.  We provide \texttt{VoigtFit} with the line-spread function (LSF) for COS/G130M Lifetime Position 4 (\citealt{2024cosi.book...17H}). \texttt{VoigtFit} convolves the Voigt profile model with the LSF to characterize the absorption. When the fitting parameters do not converge---whether because of line saturation, blending, or low S/N--- we fixed divergent line parameters such that they agree with matching parameters of similar ionization species.

For both of our sightlines, we apply a co-fit of at least two components that correspond to detected \hi~21-cm emission of the MW and the SC. To distinguish these components from each other in absorption, we usually use relatively weak lines to avoiding line saturation, like S\textsc{~ii}$\lambda \lambda 1250,1253,1259$ or P\textsc{~ii}$\lambda 1152$. We also use strong absorption lines, such as Si\textsc{~ii}$\lambda\lambda 1190,1193,1260$ and C\textsc{~ii}$\lambda 1334$, to probe for asymmetries. The UV absorption lines do not always align well with the \hi\,21-cm emission, which is likely because the \hi\ observations span a much larger angular extent than the UV observations and the resultant spectra represent an average over more gas that could have a larger variation in speeds.

For ions with multiple observed transitions, we co-fit all available lines with \texttt{VoigtFit} to reduce uncertainties for the Doppler parameters and column densities. Co-fitted transitions include at least one transition that is not saturated or weak. Additionally, if two ionization species (X and Y) are in local kinetic equilibrium, then the ratio of their Doppler widths will scale with the masses of the ions $b_X/b_Y = (m_X/m_Y)^{-1/2}$. When lines are blended, saturated, or below the 3$\sigma$ significance threshold, we superimpose the kinematic parameters corresponding to absorption lines in similar ionization states.

\begin{figure*}
\begin{center}
    \includegraphics[width=18cm]{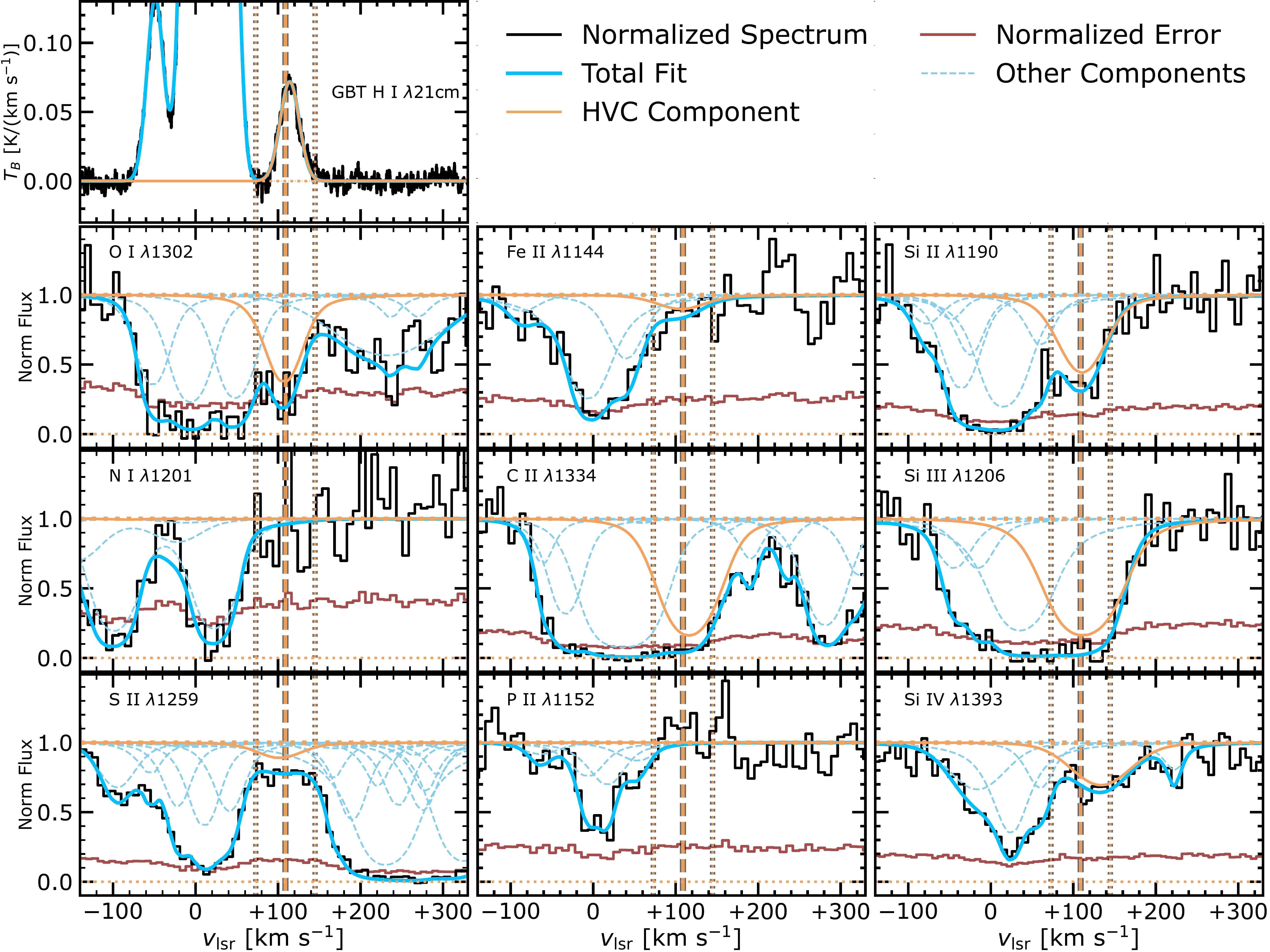}
    \caption{Sightline~D plotstack. Included are the fits for the \hi\ emission and absorption for O\textsc{~i}\,$\lambda\,1302$, S\textsc{~ii}\,$\lambda\,1259$, Si\textsc{~ii}\,$\lambda\, 1190$, Si\textsc{~iii}\,$\lambda\,1206$, and Si\textsc{~iv} $\lambda\,1393$. The black histogram corresponds to the binned normalized flux with a bin factor of 3 and the maroon histogram represents the associated flux error. The light blue curve is the total best fit to the data. The orange and blue curves are HVC and IVC component fits, respectively. The orange curve marks the \hi\ line at $v_{\rm{LSR}} = +114.7\,\kms$. The vertical orange dashed line corresponds to the \hi\, velocity centroid, with dotted vertical lines bracketing $2\sigma_v$ for the \hi\, added to and subtracted from the \hi\, velocity centroid. The absorption at $v_{\rm{LSR}} \approx -80\,\kms$ in S\textsc{~ii}$\,\lambda1259$ corresponds to blueshifted HVC absorption in Si\textsc{~ii}$\,\lambda 1260$, which we illustrate in Figure \ref{fig:J2033-0320_HVCs}.} 
    \label{fig:J2033-0320_Plostack}
\end{center}
\end{figure*}

\begin{figure*}
\begin{center}
    \includegraphics[width=18cm]{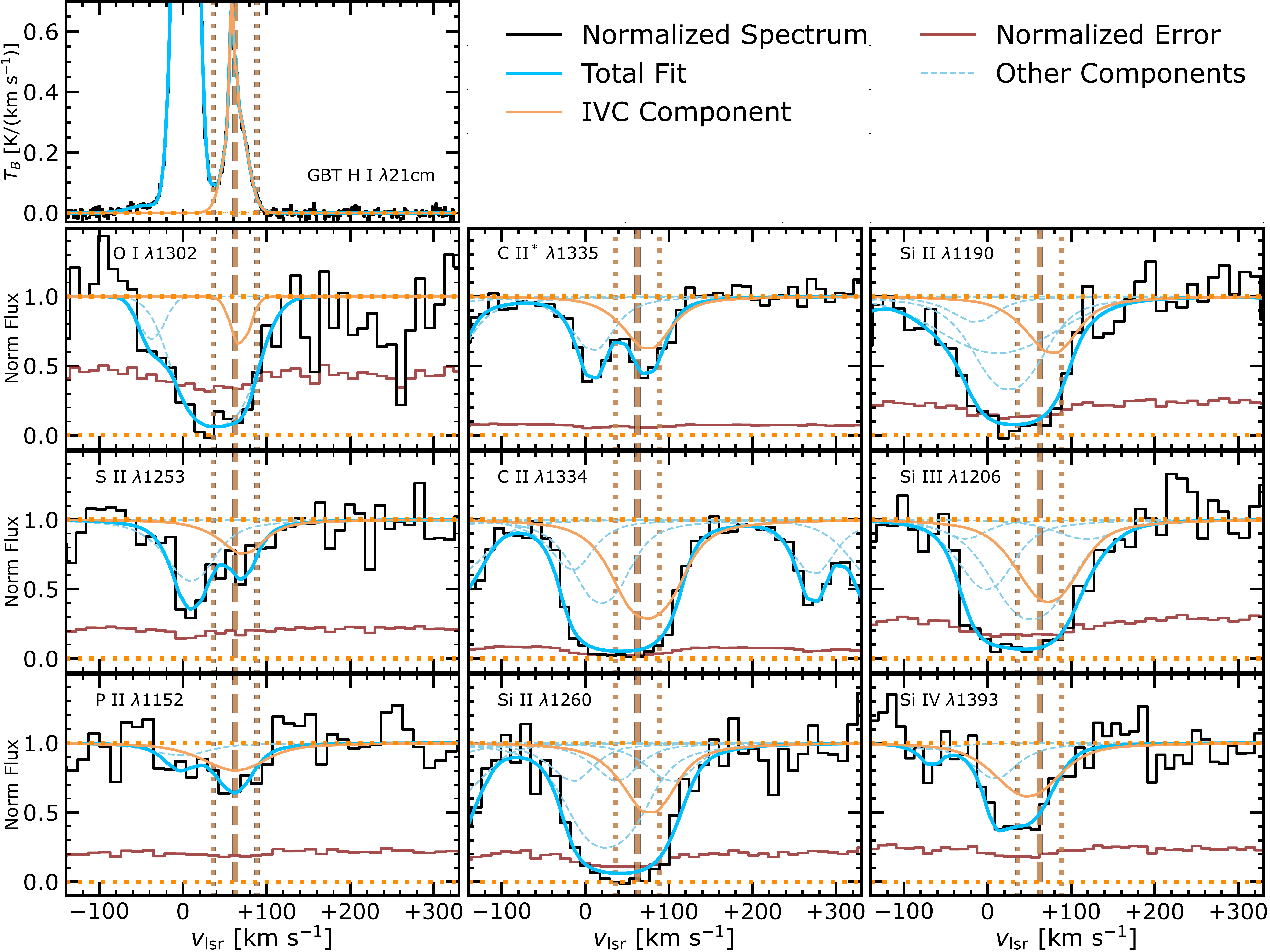}
    \caption{Sightline~E plotstack. Similarly to Figure~\ref{fig:J2033-0320_Plostack}, we plot the total fit in light blue and the IVC component in orange, which lies at $v_{ _{\rm{lsr}}} = +63.1\,\kms$ and has a width of $\rm{FWHM} = 30.1\,\kms$ as measured from the \hi\ emission.} \label{fig:J2044-0728_Plotstack}
\end{center}
\end{figure*}

For the low-ionization part of this analysis, we focus on the 
S\textsc{~ii} $\lambda\lambda\, 1250,1253,1259$, O\textsc{~i} $\lambda\, 1302$, N\textsc{~i} $\lambda\lambda\,1199.5,1200.2,1200.7$, Si\textsc{~ii} $\lambda\lambda\,1190,1193,1260,1304$, Si\textsc{~iii} $\lambda\,1206$, and Fe\textsc{~ii} $\lambda\lambda\,1143,1144$, and P\textsc{~ii} $\lambda\,1152$, C\textsc{~ii} $\lambda\,1334$, and C\textsc{~ii} $^*\lambda 1335$ lines. Although the N\textsc{~i} $\lambda\,1199,1200.2,1200.7$, and O\textsc{~i} $\lambda\,1302$ lines could provide some more insight on the volatile elements, these lines have lower signal-to-noise than the other ionic transitions due to the need to only use night-time observations; when possible, we report column density limits for them. We use the S\textsc{~ii} and P\textsc{~ii} lines to characterize the volatile elements. F16 has already completed the UV analysis for sightlines A, B, and C. Here we discuss the UV fitting for the gas along sightlines D and E below. In Section~\ref{section:discussion}, we will discuss the results for all sightlines in a global context.

\subsection{Sightline~D}
Sightline~D is positioned offset the main body of the SC and lies off of an \hi\ cloud that has presumably fragmented from this HVC (see~Figure \ref{fig:SC_map}). This sightline enables us to probe the physics at the interface of this cloudlet and the surrounding coronal gas. We detect Si\textsc{~ii} and S\textsc{~ii} absorption along this sightline that is consistent with the Milky Way at $v_{\rm{LSR}} = +42.0\pm5.7\,\kms$ and HVC absorption at $v_{\rm{LSR}} = +109.8\pm2.3\,\kms$ (see Figure \ref{fig:J2033-0320_Plostack}). This component structure is consistent with what we observe in the \hi\ 21-cm emission (see Table~\ref{table:profile_fits}). 

\begin{center}
    \begin{figure}
        \centering
        \includegraphics[width=8.6cm]{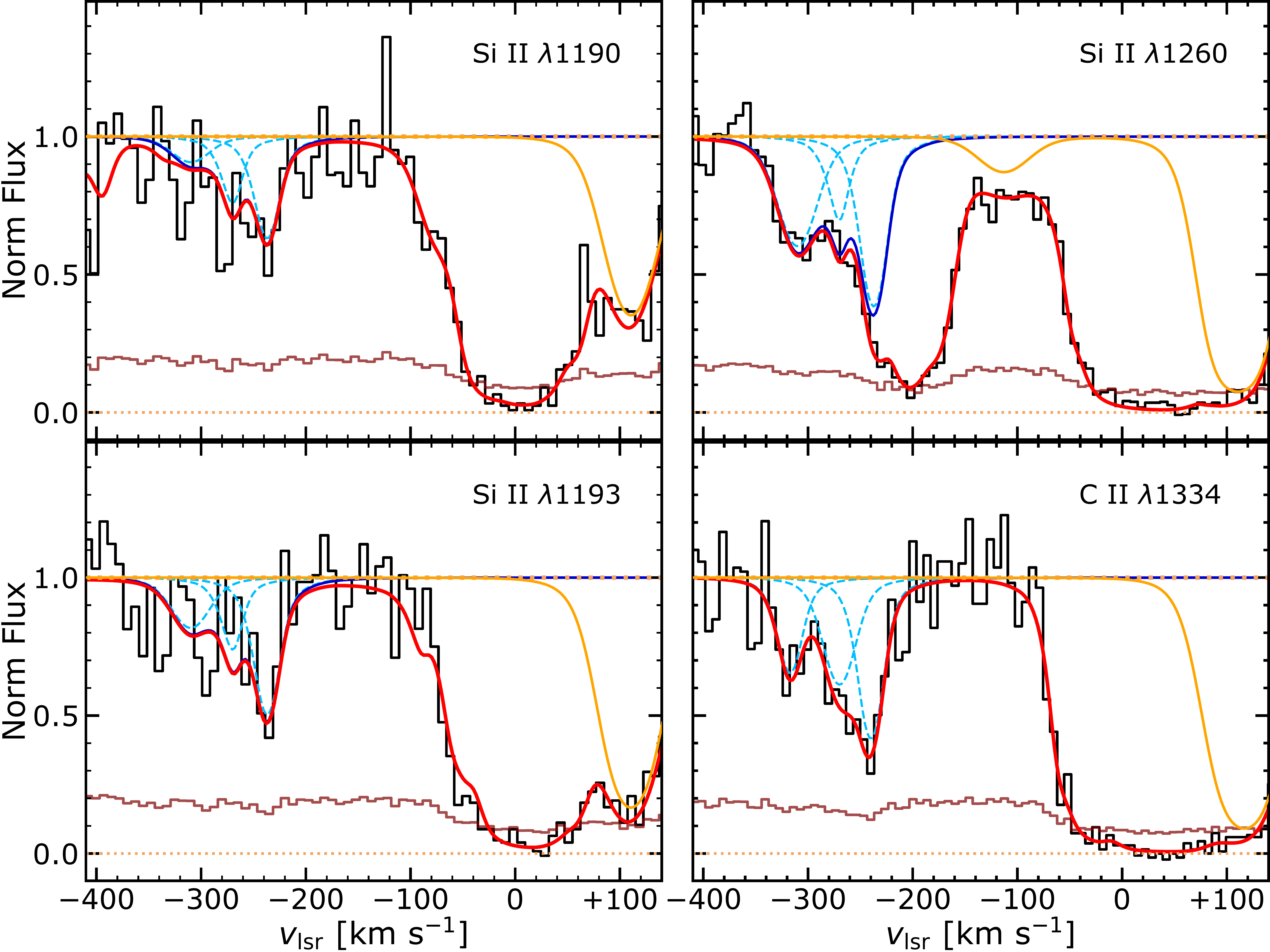}
        \caption{Spectrum of line transitions along sightline~D that are affected by blueshifted HVC components. We represent individual fits to the blueshifted Si\textsc{~ii} HVCs that span from $-315\lesssim v_{\rm{lsr}}\lesssim -235\,\kms$ with the light-blue dashed curves. We represent the combined fit of \textit{only} the blueshifted HVCs, with a dark blue solid curve. We trace the combined fit from \textit{all} absorption components with a red curve.  We also plot the SC fits in both Si\textsc{~ii} and C\textsc{~ii}$\lambda 1334$  as an orange solid curve.}
        \label{fig:J2033-0320_HVCs}
    \end{figure}
\end{center}

 Due to line blending, we co-fit all transitions of O\textsc{~i}, Si\textsc{~ii}, and S\textsc{~ii} together with blending lines.  In the Si\textsc{~ii} lines, there are HVCs with velocities $-315 \lesssim v_{\rm{lsr}} \lesssim -235\kms$; the Si\textsc{~ii}$\lambda 1260$ absorption of this HVC blends with SC and MW S\textsc{~ii}$\lambda 1259$  absorption (see Figure \ref{fig:J2033-0320_HVCs}). Unfortunately, the non-contaminated S\textsc{~ii}$\lambda \lambda 1250,\,1253$ lines are too weak to independently detect SC absorption. Despite the blending in the S\textsc{~ii}$\lambda~1259$ absorption, we use this stronger S\textsc{~ii} transition to estimate S\textsc{~ii} column density. In our Voigt-fitting regime for this sightline, we account for the following sources of blending:
\begin{itemize}
\item Weak C\textsc{~i} lines at $\lambda=1193$, $1194$, $1188$, $1260\,\AA$ that overlap with Si\textsc{~ii}$\lambda=1193,\,1190,\,1260\AA$. To independently verify the C\textsc{~i} MW absorption, we co-fit our aforementioned fainter C\textsc{~i} MW absorbers with C\textsc{~i} lines at $\lambda=1139,1157,1328\AA$. 
\item A fast, blueshifted absorber in Si\textsc{~ii}$\lambda\lambda 1260,1190$ at $-315 \lesssim v_{\rm{lsr}} \lesssim -230\,\kms$ that blends with MW's S\textsc{~ii}$\lambda 1259$ and C\textsc{~i}$\lambda1188$ absorption, respectively. 
\item An intermediate-velocity absorber at $v_{\rm{lsr}}\approx-50\,\kms$ that we detect in the Si\textsc{~ii}$\lambda\lambda 1260,1190,1193$ transitions. This absorber blends with the SC's S\textsc{~ii}\,$\lambda1259$ and the MW's S\textsc{~iii}\,$\lambda1190$ and C\textsc{~i}\,$\lambda1193$ absorption, respectively.
\item Background Ly$\alpha$ absorbers at $z = 0.07204$ and $z=0.07340$, which blend with the blueshifted Si\textsc{~ii}$\lambda 1304$ HVC absorption and Si\textsc{~ii}$\lambda 1304$ SC absorption, respectively.
\end{itemize}
We illustrate our Si\textsc{~ii} blueshifted HVC and IVC fits in Figure~\ref{fig:J2033-0320_HVCs} as blue dashed lines.

Among the high ions, we could only detect Si\textsc{~iv},  which has a centroid velocity of $v_{\rm{lsr}} = +133.2 \pm 9.9\,\kms$. This has a velocity difference with Si\textsc{~ii} of $\delta v = 23.5 \pm 10.2\,\kms$ (see Table \ref{table:profile_fits}). This velocity difference could imply disruption of the ionized gas along this sightline. The large uncertainty in the Doppler  parameter is largely attributable to background contaminating Ly$\alpha$ which obscures the Si\textsc{~iv}$\lambda 1393$ line (see Figure \ref{fig:J2033-0320_Plostack}). Given the large $b$-value and uncertainties of $b = 40.5\pm22.5$, this could indicate multiple velocity phases in the ionized gas, which could point to apparent headwinds from the MW halo stripping off SC material through ram-pressure interactions.

\subsection{Sightline~E }
This sightline~could probe the gas adjacent to the main body of the SC within a high-column density \hi\ filament (see Figure \ref{fig:SC_map}). The kinematic structures of the \hi\ gas along this sightline~are not resolved by the \textit{HST} observations as COS/G130M only has a spectral resolution of $17\lesssim v_{\rm{FWHM}}\lesssim 25\,\kms$. In the \hi\ emission, we detect a cluster of  three  components either associated with the SC or SAMW, and another component associated with the MW (see Section~\ref{subsubsection: SLE_HI_analysis} and Figure~\ref{fig:J2044-0728_Plotstack}). For the purposes of calculating the abundances of the gas along this sightline, we compare SC's UV absorption with the  sum of the three \hi\, emission components associated with the SC or SAMW. Additionally, we re-emphasize the uncertainty of the origin(s) of this \hi\ cloud and the possibility that some of the absorption and emission in these velocities could be associated with the SAMW, as discussed in Section~\ref{subsection:HI_Analysis}. 

 Along this sightline, there are well-separated absorption components matching the kinematics of the \hi\ 21-cm emission\, in P\textsc{~ii}$\,\lambda 1152$, S\textsc{~ii}$\,\lambda\lambda 1250,1253,1259$, and in C\textsc{~ii}$^*\,\lambda 1335$. We report our Voigt profile-fit values in Table~\ref{table:profile_fits}. Unfortunately, the O\textsc{~i}, C\textsc{~ii}, Si\textsc{~ii}, and Si\textsc{~iii} lines were all too saturated to reliably differentiate the absorption feature at $+50 \leq v_{\rm{lsr}} \leq +80\,\kms$ from the MW.

 In addition to the low ions, we also characterize the high ions via Si\textsc{~iv}$\,\lambda\lambda 1393,1402$ absorption. The presence of both  Si\textsc{~iv} transitions  allows us to more confidently constrain the kinematics of the ionized gas. We find  a component of Si\textsc{~iv} absorption at a velocity of $v_{\rm{LSR}}~=~+47.0~\pm~ 7.8~\,~\kms\;(v_{\rm{GSR}}~=~+168.0~\pm~7.8~\,~\kms)$.  This component is redshifted relative to the \hi\, emission and to the S\textsc{~ii} and P\textsc{~ii} absorption. In the case that the \hi\, emission along this sightline probes the SC, the lower-velocity Si\textsc{~iv} component could indicate disruption of the highly-ionized gas phase. However, we stress that this component may be blended with material from the SAMW. If the \hi\, emission along this sightline indeed probes the SC, it does so in a region of the SC where the neutral gas is more shielded than the gas along sightline~D.

\begin{deluxetable*}{c c c c c}\label{table:col_dens}
\tablecaption{Voigt profile values for this study's sightlines in addition to ionic abundances compared to neutral hydrogen.  We report 1$\sigma$ errorbars.}
\tablecolumns{5}
\tablewidth{0pt}
\tablehead{
\colhead{\textbf{Ion}} & \colhead{$v_{\rm{LSR}}$} & \colhead{$\log{ N_{\rm{X}^i}}$} & \colhead{$b$} & \colhead{[X$^i$/\hi]\tablenotemark{a}}   \\ 
  \colhead{}& \colhead{(\kms)} & \colhead{(\cmsq)} & \colhead{(\kms)} & \colhead{} 
}
\startdata
\hline \multicolumn{5}{c}{{\bf Sightline~D}} \\ 
\hline \hline \hi & $+114.7\pm 1.4$  &  $ 18.56\pm 0.25$& $24.6\pm2.3$ & $0.00$ \\
\hline
 O\textsc{~i}\tablenotemark{c,e} & $+105.4\pm 7.3$  &  $14.71 \pm 0.21$ & $17.1\pm11.9$ & $-0.54\pm 0.33\pm 0.15$  \\   
 N\textsc{~i}\tablenotemark{d} & $+110$\tablenotemark{b} & $\lesssim 14.00$ & $30$\tablenotemark{b} & $\lesssim -0.39$\\
S\textsc{~ii} & $+100.3\pm10.2$ &  $14.22\pm0.20$ & $28.3\pm19.6$ & $+0.54\pm0.33\pm 0.15$  \\ 
P\textsc{~ii}\tablenotemark{d}& $+110$\tablenotemark{b} & $\lesssim 13.15$ & $\cdots$  & $\lesssim+1.18$   \\ 
C\textsc{~ii}\tablenotemark{c} & $+110$\tablenotemark{b} & $\gtrsim 14.30$ & $30 \tablenotemark{b}$ & $\gtrsim -0.69$\\
Fe\textsc{~ii}\tablenotemark{d} & $+110$\tablenotemark{b} & $\lesssim13.61$& $\cdots$ & $\lesssim -0.45$\\
Si\textsc{~ii} & $+109.8\pm 2.3$ & $13.91\pm0.05$ & $28.2\pm2.4$ & $-0.16\pm 0.25\pm 0.15$\\
Si\textsc{~iii}\tablenotemark{c} & $+110\tablenotemark{b}$ & $\gtrsim 13.90$ & $30$\tablenotemark{b}& $\lesssim -0.17$\\
Si\textsc{~iv} & $+133.2\pm9.9$ & $13.24\pm0.13$ & $40.5\pm 22.5$ & $-0.83\pm 0.13\pm 0.15$   \\
\hline \multicolumn{5}{c}{{\bf Sightline~E}}\\ 
\hline \hline \hi & $+62.4\pm 0.4$ & $19.26\pm0.02$ & $17.3\pm0.7$ & $0.00$\\
 \hi & $+58.2\pm 0.1$ & $18.69\pm 0.02$ & $4.0\pm0.2$ & $0.00$\\
 \hi & $+64.7\pm 0.3$ & $17.69\pm 0.12$ & $1.9\pm 0.5$ & $0.00$\\
\hi\,(co-added)& $+63.0 \pm 2.4$ & $19.37\pm0.02$ & $17.8\pm0.7$ & $0.00$\\
\hline
S\textsc{~ii} & $+70.3\pm4.2$ &  $15.13\pm 0.24$ & $9.7\pm 4.1$ & $+0.64\pm0.24\pm0.15$  \\ 
P\textsc{~ii} & $+60.1\pm7.9$ & $13.54\pm0.19$ & $10$\tablenotemark{b} & $+0.76\pm0.19\pm0.15$   \\ 
C\textsc{~ii}\tablenotemark{c} &  $+75$\tablenotemark{b} & $\gtrsim 16.04$ & $20$\tablenotemark{b} & $\gtrsim +0.24$ \\
C\textsc{~ii}$^*$ &  $+75.2 \pm 2.3$ & $14.05\pm0.05$ & $19.9\pm4.2$ & $ -1.75\pm0.07\pm0.15$ \\
Fe\textsc{~ii}\tablenotemark{d,f} & $+75$\tablenotemark{b} & $\lesssim 14.74$& $\cdots$& $\lesssim -0.13$\\
Si\textsc{~ii}\tablenotemark{c} & $+75$\tablenotemark{b} & $\gtrsim 14.24$ & $20$\tablenotemark{b} & $\gtrsim-0.64$   \\
Si\textsc{~iii}\tablenotemark{c} & $+75$\tablenotemark{b} & $\gtrsim13.32$ & $20\tablenotemark{b}$ & $\gtrsim-1.56$   \\
Si\textsc{~iv} & $+47.0\pm7.8$ & $13.48\pm 0.18$& $14.2\pm 13.3$ & $-1.40\pm 0.18\pm0.15$ \\
\enddata 
\tablenotemark{a}{We add a 0.15 dex error in quadrature, due to the effects of beam dilution.}
\tablenotetext{b}{A fixed parameter in the fit have been forced to match a free fitted value for a transition of a similar ionization species with the lowest uncertainty. These values have been rounded up or down to the nearest multiple of 5\,\kms.
}
\tablenotetext{c}{A saturated line with the reported lower limit on the column density determined using Voigt profile fitting using the Python \texttt{VoigtFit}.}
\tablenotetext{d}{An absorption line below the $3\sigma$ detection limit, for which we use the AOD method to determine the upper limit. We do not report Doppler parameters for these lines.}
\tablenotetext{e}{The O\textsc{~i}$\lambda 1302$ fit is possibly saturated. For transparency, we report the \texttt{VoigtFit}-derived column density with $1\sigma$ errors in this Table. However, in the rest of the paper, we will treat the O\textsc{~i} column density as a lower limit.}
\tablenotetext{f}{This Fe\textsc{~ii}$\lambda \lambda 1143,1144$ line is too blended with background Ly$\beta$ to be discernible. We determine an upper limit from Fe\textsc{~ii}$\lambda 1142$.}
\label{table:profile_fits}
\end{deluxetable*}

\section{Photoionization Modeling}\label{section:cloudy}

While UV absorption and radio \hi\ emission spectroscopy provide  useful insight on the properties of the SC, they are limited in scope by themselves. This is because our \hi\ 21-cm spectroscopic observations can only directly account for neutral hydrogen and our UV spectroscopy only provide insight into the column densities of \textit{certain} ionization species within the SC. We used radio observations to constrain the hydrogen because the Lyman absorption of the MW is too strong such that any signal from diffuse HVCs cannot be resolved. Therefore, the [X$^i$/H\textsc{~i}] relative abundances we find in Section~\ref{section:specanalysis} are only a first-order estimate of the true abundances. 

To correct for the presence of ions hidden in unseen ionization stages, we implement \cloudy\,23.00 photoionization models \citep{2023RMxAA..59..327C}. For the Milky Way's radiation field, we use the \citet{2005ApJ...630..332F} 3~dimensional model for the Milky Way to estimate hydrogen ionizing flux $\phi$ along each of our sightlines at an assumed distance of $d_\odot = 12.4\pm 1.3\kpc$. Additionally, we use the \citet{2019MNRAS.484.4174K} model for the extragalactic UV background radiation field; however, since the SC only lies a few kpc below the Galactic plane, the Milky Way's emission dominates the ionizing radiation it experiences. For simplicity, we assume collisional ionization is negligible even though collisional interactions with the surrounding halo medium may be important for disrupting the HVC's morphology. Here we outline our exploration of the chemical depletion patterns in the SC using only \cloudy\ radiative transfer models in conjunction with our GBT and \textit{HST} observations. 

\subsection{Optimized Models}\label{subsection:OPT}

The Si\textsc{~iii} absorption along all of our sightlines is saturated, which impacts how well we are able to constrain the ICs. Additionally, due to the low \hi\ column density along sightline~D, the limit on our Si\textsc{~iii}/Si\textsc{~ii} ratio is insufficient to meaningfully bound our ICs as it is highly sensitivity to variations in $N_{\rm{\hi}}$. However, we caution that the gas is likely in a multiphased state along the explored sightlines as they trace the halo-HVC interface and these radiative transfer models assume that the gas is well mixed. Therefore, we additionally use \cloudy's optimize model functionality to constrain parameter space to bracket the ICs. 

We ran optimized \cloudy\ photoionization models for $(2\times N)$-dimensional parameter sets of metallicities ($Z$) and the hydrogen volume density ($n_{\rm{H}}$). We specifically bound these models so that they are consistent with the observed UV-based ionic column densities of volatile elements and radio-based \hi\ column density using a $\chi^2$ minimization. 

For sightlines that have S\textsc{~ii} column densities with large uncertainties, we ran three sets of optimized models anchored at $N_{\rm{S\textsc{~ii},\,input}}=\langle N_{\rm{S\textsc{~ii},\,obs}}\rangle$, $N_{\rm{S\textsc{~ii},\,input}}=\langle N_{\rm{S\textsc{~ii},\,obs}}\rangle+\sigma_{N_{\rm{S\textsc{~ii}}}}$, and $N_{\rm{S\textsc{~ii},\,input}}=\langle N_{\rm{S\textsc{~ii},\,obs}}\rangle-\sigma_{N_{\rm{S\textsc{~ii}}}}$. We report the median value of the  IC values from consistent models, as well as, the upper and lower bounds when we ran three sets of simulations in Table~\ref{table: ICs_and_dust}.
\section{Abundances} \label{section:abundances}

Gas-phase abundances provide clues into the likely origin and history of the SC. For instance, high metal content and dust depletion would likely be indicative of either Milky Way origin or a past interaction with the plane of the Galaxy.

Past measurements and interpretations of the SC's metallicity are in disagreement, pointing to the need for more information on the chemical distribution at different regions of this HVC (\citealt{2016ApJ...816L..11F}; \citealt{2009ApJ...703.1832H}). Additionally, information on the abundance patterns of this HVC can provide insight into its history.

To estimate the total column densities of volatile elements, we correct the observed ionic column densities for ionization using the ionization corrections listed in Table~\ref{table: ICs_and_dust} (see Section~\ref{section:cloudy}), which are related by:
\begin{equation}
    \text{IC(X}^i) = \log\left(\frac{\xi_{\rm{\hi}}}{\xi_{\rm{X}^i}}\right)
\end{equation}
The $\xi_{\rm{X}^i}$ represents the ratio of an ionic column density ($N_{\text{X}^i}$) to the total column density ($N_{\rm{X,tot}}$) of element~X:
\begin{equation}
    \xi_{\rm{X}^i} = \frac{N_{\text{X}^i}}{N_{\rm{X,tot}}}
\end{equation}
Rewriting this expression in terms of model parameters and observed parameters, it becomes:
\begin{equation*}
   \text{IC(X}^i) = \text{[X/H]}_{\text{model}} - \text{[X}^i\text{/H\textsc{~i}]}_{\text{observed}},
\end{equation*}
Here, [X/H]$_{\rm{model}}$ is gas-phase abundance of element~X that we constrained using \cloudy\, modeling. The ionic gas-phase abundance ([X$^i$/\hi]), which is not corrected for ionization, is:
\begin{equation}
    \text{[X}^i\text{/H\textsc{~i}]}_{\text{obs}} = \log\left(\frac{N_{\rm{X}^i}}{N_{\rm{\hi}}}\right)_{\rm{obs}} - \log\left(\frac{\rm X}{\rm H}\right)_\odot.
\end{equation}
In addition to gas-phase abundances, we can also explore gas-phase depletions by choosing a volatile reference element that is undepleted in the warm-ionized medium. Since our O\textsc{~i}$\lambda 1302$ lines are saturated, we choose S\textsc{~ii} as our reference ion. We can obtain the gas-phase depletion $\delta_\text{S}(\rm{X})$ of an element X by comparing it to the gas-phase depletion of sulfur as follows:
\begin{equation}
    \delta_{\rm{S}}(\rm X) \equiv  [X/S] \equiv [X/H] - [S/H]
\end{equation}

In order to explore parameter space with \texttt{optimize} models, we must provide reasonable bounds on our input parameters. Applying the 3$\sigma$ bounds on the metallicity results of the F16 study, we expect the metallicity of the SC to be within the range of $-1.3 \lesssim \log(Z/Z_\odot) \lesssim +0.5$. In addition, we can provide rough constraints on the model volume density. We choose $\log(n_{ _{\rm{H}}}/\rm{cm}^{-3}) \lesssim -1.0$ as our upper bound, because the mean electron density of the SC is estimated to be $\log(n_e /\rm{cm}^{-3}) \approx -1.13$ \citep{2013ApJ...777...55H}. We assume the gas along each of these sightlines is highly ionized and that it is mixing with the Galactic halo. Therefore, we assume that $n_{\rm{H}} \approx n_e$ along this sightline, which is likely lower than the average of the entire SC due to lower \hi\, column densities than in the main body. We also estimate a lower limit for our hydrogen densities using the largest value of the line-of-sight ionized gas length  $L_{\rm{H\textsc{~ii}}}=1000\pc$ from \cite{2013ApJ...777...55H} to find a lower limit of the hydrogen density of $\log(n_{ _{\rm{H}}}/\rm{cm}^{-3}) \gtrsim -2.83$. For the gas along each sightline, we therefore use a lower bound of $\log(n_{ _{\rm{H}}}/\rm{cm}^{-3}) \gtrsim -3.0$.

\begin{center}
\begin{deluxetable*}{c c c c c c c c c}\label{table:weighted_Zs}
\tablecaption{A cumulative table including the metallicities, volume densities, \hi\, column densities, ionizing photon fluxes, LSR velocities, and Galactic coordinates of all SC QSO sightlines that have been studied. We report 1$\sigma$ errorbars.}
\tablecolumns{9}
\tablewidth{0pt}
\tablehead{
\colhead{Sightline} \vspace{-0.2cm} & \colhead{Source} & \colhead{$\ell_{\rm{gal}}$} & \colhead{$b_{\rm{gal}}$}& \colhead{$\log\phi$} & \colhead{$v_{\rm{LSR}}$}&  \colhead{$\log N_{\rm{H\textsc{~i}}}$} &\colhead{[S/H]\tablenotemark{b}} & \colhead{$\log n_{ _{\rm{H}}}$\tablenotemark{c}}  \\& & \colhead{$(^\circ)$}  & \colhead{$(^\circ)$} &  \colhead{$\left(\cmsq\,\rm{s}^{-1}\right)$}&\colhead{$\left(\kms\right)$} & \colhead{$\left(\cmsq\right)$}  & &\colhead{$\left(\rm{cm}^{-3}\right)$}
}
\startdata
A & F16 & $+49.72$ & $-22.88$ & $5.78$  & $+81.8\pm 1.6$ & $18.81\pm0.04$  &$-0.14\pm 0.20$ & $-1.50\pm 0.10$ \\
B & F16 \& This work& $+57.04$ & $-28.01$ & 5.78  &  $+49.7\pm 2.6$  & $18.37\pm0.09$  &$-0.35^{+0.36}_{-0.47}$& $-1.75^{+0.15}_{-0.27}$\\
C & F16 & $+58.09$ & $-35.01$ &  5.78 &$+58.1\pm 0.4$ & $19.32\pm0.02$ & $-0.58\pm 0.25$ & $-1.70\pm 0.10$ \\
 D & This work & $+42.04$  &  $-24.18$ & 5.82 &$+114.7\pm 1.4$& $18.56\pm 0.25$  & \shSLA &$-1.62^{+0.10}_{-0.26}$ \\    
E\tablenotemark{a} & This work & $+38.79$& $-28.62$ &6.03 &$+63.0\pm2.4$& $19.37\pm0.02$ &  $+0.28^{+0.25}_{-0.26}$ & $-1.55^{+0.30}_{-0.44}$  \\ 
\enddata
\tablenotetext{a}{As discussed in Section \ref{subsection:HI_Analysis}, we cannot be certain that the gas along this sightline~Belongs to the SC. We include the analysis in this table for completeness' sake.}
\tablenotetext{b}{In addition to the statistical error, we also propagate an beam-smearing error of 0.15 dex, equivalent to that used in F16.}
\tablenotetext{c}{For Sightlines A and C we derive $n_{\rm{H}}$ values from the ionization parameter ($U$) values reported by F16.}
\end{deluxetable*}
\end{center}

\begin{center}
\begin{deluxetable*}{c c c c c c c}
\label{table: ICs_and_dust}
\tablecaption{Cloudy-derived ionization characteristics, assuming an incident hydrogen ionizing flux from \cite{2005ApJ...630..332F}. We graphically display these values in Figures~\ref{fig:J2033-0320_cloudy_figs}~and~\ref{fig:J2044-0758_cloudy_figs}.
}
\tablecolumns{7}
\tablewidth{0pt}
\tablehead{
\colhead{\textbf{Ion} (X$^i$)} \vspace{-0.2cm} & \colhead{[X$^i$/\hi]} & \colhead{[X$^i$/S\textsc{~ii}]} & \colhead{IC(X$^i$)} &  IC$_{\rm{S\textsc{~ii}}}$(X$^i$) &\colhead{[X/H]\tablenotemark{a}} & \colhead{[X/S]}}
\startdata
\multicolumn{7}{c}{\textbf{Sightline~D $-$ HVC}}\\ 
\hline
O\textsc{~i}\tablenotemark{b} & $\gtrsim -0.54$  & $\gtrsim -1.30$ & $ -0.03^{+0.02}_{-0.01}$& $+0.97^{+0.15}_{-0.17}$ &$\gtrsim -0.57$  & $\gtrsim -0.33$  \\ 
N\textsc{~i}\tablenotemark{c} & $\lesssim -0.39$ & $\lesssim -0.93$ & $+0.27^{+0.05}_{-0.08}$ &$+1.27^{+0.20}_{-0.26}$ & $\lesssim +0.12$ & $\lesssim +0.34$ \\
P\textsc{~ii}\tablenotemark{c} & $\lesssim +1.18$ & $\lesssim +0.64$ & $-1.32^{+0.16}_{-0.13}$ & $-0.32\pm0.05$& $\lesssim -0.14$ & $\lesssim +0.32$\\
S\textsc{~ii} & $+0.54\pm0.33$ & $0.00$ & $-1.00^{+0.18}_{-0.16}$ & $0.00$& \shSLA & $0.00$\\
Si\textsc{~ii} & $-0.16 \pm 0.25$ & $-0.70\pm 0.23$& $-1.03^{+0.23}_{-0.25}$ &$-0.02^{+0.05}_{-0.11}$ & \sihSLA & \sisSLA\\
Fe\textsc{~ii}\tablenotemark{c} & $\lesssim-0.45$ & $\lesssim -0.99$& $-0.57^{+0.17}_{-0.24}$ & $+0.42^{+0.05}_{-0.12}$ & $\lesssim -1.02$ & $\lesssim -0.57$ \\
\hline 
\multicolumn{7}{c}{\textbf{Sightline~E $-$ IVC}} \\ \hline
S\textsc{~ii} & $+0.64\pm0.24$ & $0.00$ &  $-0.36^{+0.06}_{-0.09}$ & 0.00& \shSLB & $0.00$ \\
P\textsc{~ii} & $+0.76\pm 0.19$ &$+0.12\pm0.31$ & $-0.54^{+0.11}_{-0.19}$ & 
 $-0.18^{+0.06}_{-0.10}$ & $+0.22^{+0.20}_{-0.21}$ & $-0.06^{+0.31}_{-0.32}$ \\
C\textsc{~ii}\tablenotemark{b} & $\gtrsim +0.24$ & $\gtrsim -0.40$& $-0.43^{+0.08}_{-0.09}$ & $-0.07^{+0.02}_{-0.05}$ & $\gtrsim -0.19$ & $\gtrsim -0.47$\\
Si\textsc{~ii}\tablenotemark{b} & $\gtrsim -0.64$ & $\gtrsim -1.28$ & $-0.57^{+0.14}_{-0.19}$ & $-0.21^{+0.08}_{-0.10}$ & $\gtrsim -1.21$ & $\gtrsim -1.49$\\
Fe\textsc{~ii}\tablenotemark{c} & $\lesssim -0.13$ & $\lesssim -0.77$ & $-0.31^{+0.04}_{-0.02}$ & $+0.05^{+0.07}_{-0.01}$ & $\lesssim -0.44$ &$\lesssim -0.72$ \\
\enddata
 \tablenotetext{a}{We report our metallicity measurements without beam-smearing in this table, but we caution the reader that there is an additional 0.15 dex \hi\, beam-smearing uncertainty. Additionally, for ions which have upper and lower limits to their column densities, we report the metallicity bound as the [X$^i$/\hi] limit added to the upper and lower limits of ICs, respectively.}
 \tablenotetext{b}{
For these ions, the absorption profile is saturated. We report our Voigt profile-derived column density value as a lower limit. For O\textsc{~i}, we report our Voigt profile-derived column density value of $\log(N_{\rm{O\textsc{~i}}}/\text{\cmsq}) = 14.70$ from Table \ref{table:profile_fits} as the lower limit on column density.}
\tablenotetext{c}{For these ions, there was no detectable absorption. We therefore take the 3$\sigma$ detection limit derived from AOD as the upper limit on column density (see~Section \ref{section:specanalysis}).}
\end{deluxetable*}
\end{center}
\subsection{Sightline~B}\label{subsection:pg2112_opt}
F16 has already studied the absorption lines along Sightline~B. Unfortunately, there were no other undepleted elements for which we could constrain our \texttt{optimize} models. Additionally, their \hi\ column density value is over a factor of 2 larger than we measure with our new, pointed observations. Therefore, it is necessary for us to re-analyze the \cloudy\, models at our newly determined \hi\ column density. We find that IC(S\textsc{~ii})~=~$-1.15^{+0.04}_{-0.30}$, giving us a metallicity of [S/H]~=~$-0.35^{+0.36}_{-0.47}$. We also find a hydrogen volume density of $\log(n_{\rm{H}}/\rm{cm}^{-3}) = -1.75^{+0.15}_{-0.27}$, which agrees with the hydrogen densities along our other sightlines. This is within 1$\sigma$ agreement to F16, who adopt a volume density of $\log(n_{\rm{H}}/\rm{cm}^{-3}) = -1.70$ and find a metallicity of [S/H]~$=~-0.09\pm 0.33$. 

\subsection{ Sightline~D} \label{subsubsection:j203335_hvc_opt}
We found results consistent with previous studies for the Sightline~D HVC. Due to the large uncertainty  in \hi\, column density, we first ran \cloudy\, grid models at $\log(N_{\rm{\hi}}/\cmsq) = 18.32, 18.56, 18.81$ at a model metallicity $\log(Z/Z_\odot) = -0.3$, equivalent to the average metallicity determined by F16. To find the densities over which to run, we applied masks based on our observed silicon ionization ratio $\log(N_{\rm{Si\textsc{~iii}}}/N_{\rm{Si\textsc{~ii}}}) \gtrsim -0.01$ (see Table \ref{table:profile_fits}). For models run at $\log(N_{\rm{\hi}}/\cmsq) = 18.32$, our density was limited by $\log(n_{\rm{H}}/\rm{cm}^{-3}) \lesssim -1.48$; for models run at $\log(N_{\rm{\hi}}/\cmsq) = 18.56$, our density was limited by $\log(n_{\rm{H}}/\rm{cm}^{-3}) \lesssim -1.29$; and for models run at $\log(N_{\rm{\hi}}/\cmsq) = 18.81$, our density was limited by $\log(n_{\rm{H}}/\rm{cm}^{-3}) \lesssim -1.20$. To find the best-fit model for this sightline, we take the median of the model outputs when exploring parameter space through \hi\, column density and S\textsc{~ii} column density.

For the \textsc{~Cloudy} output parameters, we find a model density of $\log(n_{\rm{H}}/\rm{cm}^{-3}) = -1.62^{+0.10}_{-0.26}$, corresponding to an ionization parameter of $\log U = -3.19^{+0.26}_{-0.10}$, similar to those found by F16 in the SC. The ionization corrections of Si\textsc{~ii} and S\textsc{~ii} will be roughly equal to each other under most photoionization conditions in the low $N_{\hi}$ regime \citep{2020ApJ...892...19H}. Therefore it is expected that the relative ionization corrections between S\textsc{~ii} and Si\textsc{~ii} will be small. This strengthens the case for using [Si/S] as a measure of silicon gas-phase depletion, as the ionization correction will only make a small contribution to [Si/S].

\begin{figure}

\begin{center}

    \includegraphics[width=8.5cm]{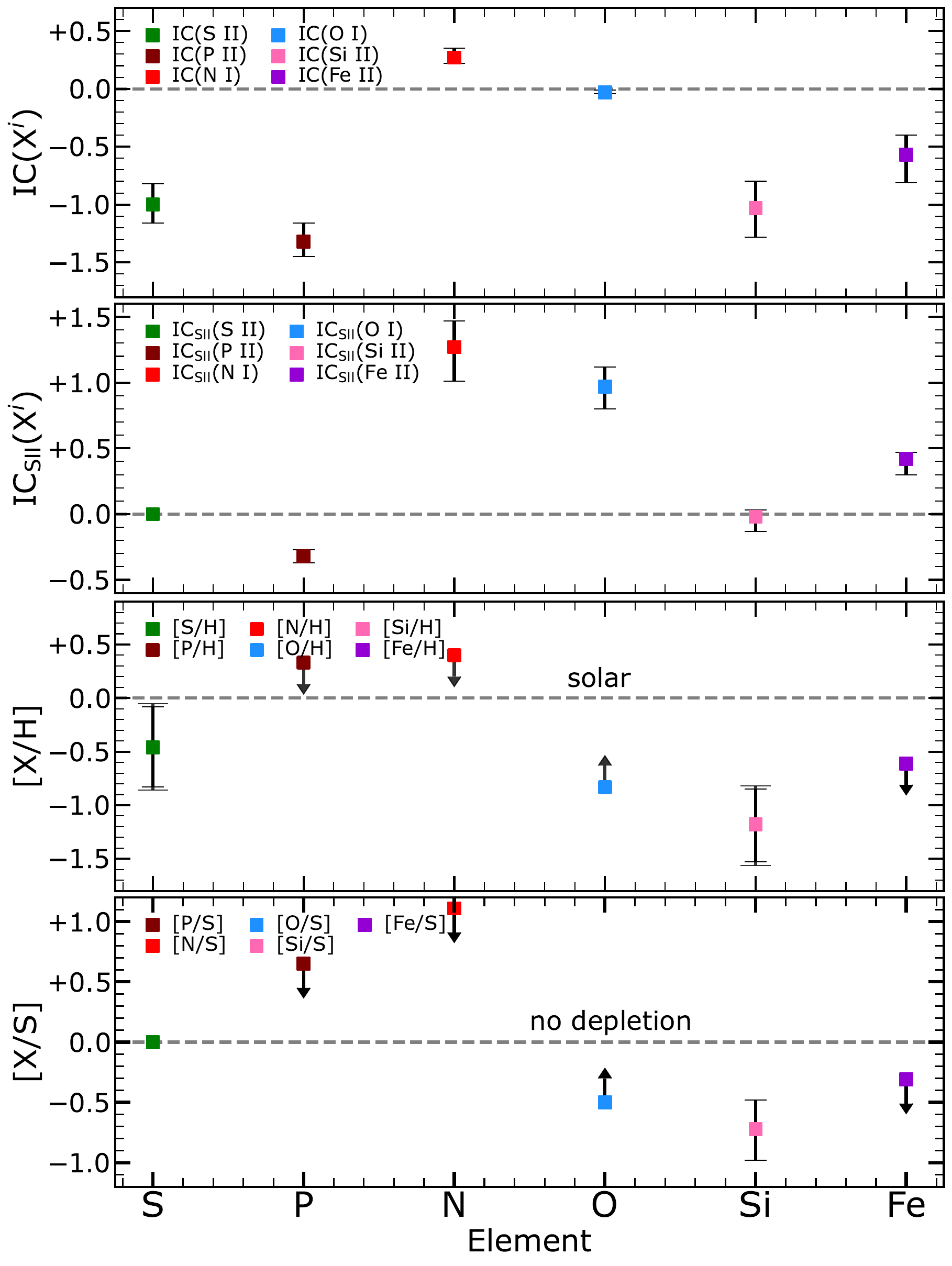}
    \end{center}

    \caption{(First panel) The ionization corrections for various ions in the SC HVC component along sightline~D.  
    (Second panel) Ionization corrections relative to S\textsc{~ii} for various ions.
    (Third panel) Gas-phase elemental abundances relative to hydrogen. The black errorbars represent the statistical error from the UV column densities. The gray errorbars represent the systematic beam-smearing error of 0.15~dex added in quadrature with the statistical error.
    Upper limits and lower limits are characterized by a down arrow and up arrow, respectively. 
    (Fourth panel) Gas-phase depletion relative to sulfur. Since \hi\ is not directly used in these calculations, we only plot the statistical errors. Values below $\rm{[X/S]} < 0\,\rm{dex}$ correspond to gas-phase depletion. All values displayed in this figure are listed in Table~\ref{table: ICs_and_dust}.} 
    \label{fig:J2033-0320_cloudy_figs}
\end{figure}

With \texttt{optimize} models, we find a sulfur abundance of [S/H]~$=$~\shSLA \,and a silicon gas-phase depletion of [Si/S]$~=~$\sisSLA, a $3\sigma$ measurement. If we compare the silicon gas-phase depletion to oxygen, it becomes much clearer that we have gas-phase depletion. If the O\textsc{~i}$\lambda 1302$ line is saturated, the [Si/O] measurement can be a reliable upper limit on silicon gas-phase depletion. If not, then we have relative gas-phase abundances of [Si/O]~$=-0.61^{+0.32}_{-0.34}$ and [S/O]~$=+0.11\pm0.35$. If we compare both silicon and sulfur to oxygen, the depletion strength factor, as discussed in \cite{2009ApJ...700.1299J}, becomes $F_* = 0.33\pm0.16$. This could be indicative of the presence of dust within the SC.  Since each IC is model-dependent, we add the maximum relative IC of Si\textsc{~ii} compared to O\textsc{~i}. Since our O\textsc{~i} column density is a lower limit, we add 3 standard deviations of the Si\textsc{~ii} column density error  added in quadrature to the Si\textsc{~ii} IC error  to find that our upper limit on the relative gas-phase depletion of silicon is $\text{[Si/O]}_{3\sigma}~ \lesssim$~\sioSLA. Even with the blending issues in the S\textsc{~ii}$\lambda 1259$ line, we still have good confidence in a sub-solar upper bound on the silicon gas-phase depletion. Additionally, lower limits on [C/H] and [O/H] and  the upper limits on [N/H] are in agreement with our [S/H] measurement (see Figure~\ref{fig:J2033-0320_cloudy_figs}).

\subsection{Sightline~E }\label{subsubsection:j204402_opt}
 Unfortunately, since the Si\textsc{~ii} is saturated, we can only constrain our models with the S\textsc{~ii} and P\textsc{~ii} column densities, and using limits on the C\textsc{~ii} column density. We fortunately are able to provide confident bounds on C\textsc{~ii} column density by adopting the kinematics of the C\textsc{~ii}$^*\lambda 1335$ line (see Section~\ref{section:specanalysis}). We find a supersolar sulfur abundance of $[{\rm S/H}]=+0.28^{+0.25}_{-0.26}$ and phosphorus abundance of $[{\rm P/H}]=+0.22^{+0.20}_{-0.21}$. If this IVC is indeed part of the SC, this would be the highest metallicity yet detected in this HVC. In conjunction with the detection of dust containing silicon, this would additionally be indicative of a Galactic origin or past Galactic interaction. 
 
 As we discussed in Section \ref{subsection:HI_Analysis}, we believe that, due to the Galactic longitude of the sightline~At $\ell = +38.79^\circ$, this gas could instead be correlated with the expected velocities of SAMW material at a distance of $d_\odot \approx 7\kpc.$ If we leave the incident hydrogen ionizing flux as a free parameter in our \texttt{optimize} models, we find $\phi \approx 10^{6.22}\,\cmsq\,\rm{s}^{-1}$ as the best fit, which is consistent with a closer distance to the Galactic center than the SC. 
 
\begin{figure}

\begin{center}

    \includegraphics[width=8.5cm]{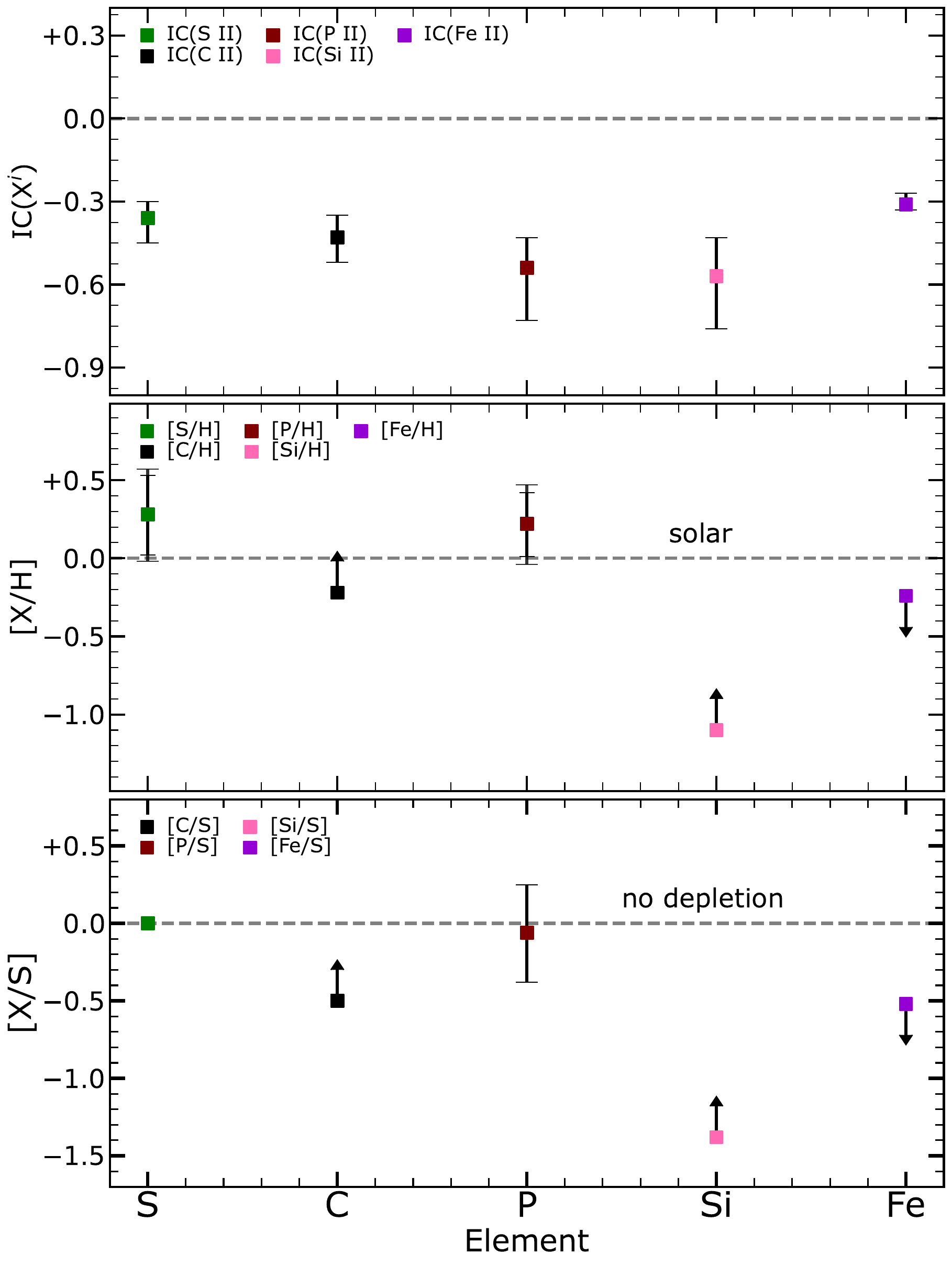}
    \end{center}

    \caption{We plot the same parameters as in Figure~\ref{fig:J2033-0320_cloudy_figs} for Sightline E, except we do not list the parameter IC$_{\rm{S\textsc{~ii}}}(\rm{X}^i)$, as S\textsc{~ii} IC is less sensitive to changes in \hi\, column density for $N_{\rm{\hi}} \gtrsim 10^{19.0}\,\cmsq$. We disseminate all values determined in Table \ref{table: ICs_and_dust}.}
    \label{fig:J2044-0758_cloudy_figs}
\end{figure}

\subsection{Results}
We determine silicon gas-phase depletion ratios of [Si/S]~=~\sisSLA\, and $\text{[Si/O]}_{3\sigma}~ \lesssim$~\sioSLA\, in the high-velocity gas along sightline~D. This is moderately strong evidence of dust containing silicon within the SC. We also find a subsolar [Fe/S] $\lesssim -0.57$ along this sightline. Iron is an Fe-peak element that is primarily deposited in the ISM through Type Ia supernovae \citep{2018A&A...616A..18E}. Whereas, $\alpha$ elements are primarily deposited in the ISM through Type II supernovae \citep{2018A&A...616A..18E}. Iron and $\alpha$ elements are not evenly distributed throughout the Galaxy \citep{2020AJ....160..120J}. Our subsolar [Fe/S] gas-phase measurement is consistent with iron dust depletion, but not necessarily indicative of iron dust depletion.  We also determine a sulfur abundance along sightline~D of ${\rm [S/H]}=$\shSLA, giving a metallicity of $Z = 0.34^{+0.49}_{-0.20}Z_\odot$, in agreement with past measurements of the SC's metallicity (F16; \citealt{2009ApJ...703.1832H}).

As it is highly likely that the high-velocity gas along sightline D is probing the SC, we must consider  our results in context with F16. F16 argued that there was some evidence for a metallicity gradient. In including our new [S/H] measurement along sightline~D, we find that any metallicity gradient is either weak or that our measurements are not sensitive enough to detect such a gradient. The high uncertainties of the SC metallicity along sightline B, and possible foreground contamination along sightline E complicate matters. We still cannot confidently confirm nor rule out metallicity gradients or metallicity mixing between the SC and the Milky Way CGM. This should be an avenue of exploration in future studies, as well as the dynamics of dust grains within the SC.

We additionally find high volatile gas-phase abundances  along Sightline~E, which  lies  near the main body of the SC, with metallicities of [S/H]=\shSLB~$\pm0.15$ and [P/H]=$+0.22~^{+0.20}_{-0.21}\pm0.15$. The gas along sightline~E has a higher density by a factor of about 2 and a higher metallicity by a factor of about 4 than the high-velocity gas along sightline~D. Furthermore, the \hi\ line widths along sightline~E are much smaller than those along sightline~D (see Table~\ref{table:profile_fits}). These differences could signify that the gas along Sightline~E is better able to cool itself, if the absorbers and \hi\, along this sightline all probe the SC. Despite the high metallicity, \hi\, at $v_{\rm{LSR}} = +63.7~\kms$ has a cometary shape consistent with the main body of the SC (see Figure \ref{fig:SC_vel_chan_maps}, left panel). However, we cannot rule out foreground contamination by the SAMW, as the velocities we probed are similar to those expected for the SAMW at $+35^\circ \lesssim \ell \lesssim +45^\circ$ (see Figure 3 in \citealt{Vall_e_2017}).

Along with metallicity, dust can be utilized as a tracer of the origins of HVCs (e.g., \citealt{2023ApJ...946L..48F}). Dust is primarily produced and destroyed within supernovae and giant stars within galaxies (\citealt{1981ASSL...88..317D}). Additionally, dust is believed to photodissociate in the Galactic halo, where it is unprotected from the Milky Way's ionizing radiation field (\citealt{1999ApJ...512..636W}). Therefore, the potential presence of dust within an HVC implies at least some of material comes from a galaxy. 

\section{Discussion}\label{section:discussion}
 There is a growing literature where the presence of dust in HVC is inferred from measurements of gas-phase depletion by using UV absorption-line spectroscopy. Most recently, \cite{2023ApJ...944...65C} and \cite{fox2023detection} discovered small gas-phase silicon depletions of [Si/O]$~=~-0.33~\pm~0.14$ within a MW HVC near the Galactic center, toward the star HD 156359; and [Si/S]~=~$-0.29\pm0.05$ within Complex C, respectively. \citet{2013ApJ...772..110F} and \citet{2022MNRAS.511.1714T} have measured the largest gas-phase depletions to date in the high-metallicity Magellanic Stream filament, with $[\text{Si/S}] = \delta\rm(Si) \approx -0.6$, and for a MW HVC along the H1821+643 sightline, with $-0.9\lesssim \delta(\rm{Si}) \lesssim -0.2$, respectively. Our silicon depletion of [Si/S]~=~\sisSLA\, is among the strongest gas-phase depletions that have been measured by this method. This points to at least some of the SC gas having a galactic origin.

 Our results in context with the metallicity distribution of the SC, determined by \cite{2009ApJ...703.1832H} and F16, beg the question: Why is there so much variation in the chemical profile of this HVC? As the SC travels toward the Galactic disk, it is interacting with its environment. If the SC is mixing with surrounding cornal medium, its metallicity could be altered in an inhomogeneous way. Unfortunately, there is disagreement with whether or not the metallicity of the MW's halo is low or high that stems from a tension between photometric metallicities of halo stars and spectroscopic gas-phase metallicities of ionized gas. The low photometric metallicities of halo stars indicate that the halo should have a metallicity around $-2.5\lesssim \log(Z/Z_\odot) \lesssim -1.5$ (e.g., \citealt{2020MNRAS.492.4986Y}). However, measurements of O~\textsc{~vii} and O\textsc{~viii} emission and models of the CGM suggest that the halo gas could have a metallicity as high as $\log(Z/Z_\odot) \approx -0.30$ ($Z \approx 0.50\,Z_\odot$; \citealt{2015ApJ...800...14M}).

It is also possible that much of the SC's dust has been mixed with the halo or the surrounding coronal medium. The gas along Sightline~D, where we observe gas-phase depletion, lies on the edge of an \hi\, cloud fragment (see Figure \ref{fig:SC_vel_chan_maps}). The following measurements imply that the SC is actively mixing with the surrounding halo gas:
\begin{enumerate}
    \item The low \hi\, column density of \newline $\log(N_{\rm{\hi}}/\cmsq) = 18.56 \pm 0.25$ (see Table \ref{table:col_dens}). This is an order of magnitude smaller than typical column densities along the main body of the SC at $\log(N_{\rm{\hi}}/\cmsq) \gtrsim 19.5$ (\citealt{2008ApJ...679L..21L}, \citealt{2023ApJ...943...55L}).
    \item The large kinematic width of the absorbers with \newline$b_{\rm{Si\textsc{~ii}}} = 28.2\pm2.4\,\kms$. The similarity of the Si\textsc{~ii} Doppler parameter to that of \hi\, is consistent with non-thermal broadening along this sightline (see Table \ref{table:col_dens}).
    \item The high ionization fraction of \newline $\chi_{\rm{H\textsc{~ii}}} = n_{\rm{H\textsc{ii}}}/n_{\rm{H}} = 96.8^{+0.72}_{-1.28}\%$. This high ionization fraction is consistent with the medium surrounding the SC (see Section \ref{subsubsection:j203335_hvc_opt}).
\end{enumerate} \cite{1996ApJ...457..211S} found that the dust depletion of MW halo clouds along HD 116852 is roughly [Si/S]$~=~-0.21$ at $z = -1.3\kpc$ and $d_\odot = 4.8\kpc$. If their measurements are representative of the dust content within the MW halo, then this mixing could dilute the dust within the SC, as the dust content of the SC is significantly higher than that of the surrounding CGM.

 UV-based gas-phase depletion  measurements\, are not the only way to detect dust. The primary method through which dust is detected in HVCs is FIR emission (e.g., \citealt{2009ApJ...692..827P}). However, there has been considerable difficulty detecting FIR-emitting dust in HVCs beyond the few detections that exist. Recently, \cite{2024ApJ...966...76M} searched for FIR-emitting molecular gas and dust from the main body of the SC. Although they were not able to detect either, they did provide upper bounds on the OH column density and FIR dust emission. From the lack of detectable dust emission, \citet{2024ApJ...966...76M} argue that the dust in the SC is at least a factor of three lower than that of the MW disk, which is known to be $\log {\rm D/G}_{\rm{MW}} \approx -2.13$ \citep{2007ApJ...663..866D}. This implies that the constraints on the SC dust fraction is $\log {\rm D/G}_{\rm{SC}} \lesssim -2.61$.  We can calculate our measured dust-to-gas ratio via the following relationship:
\begin{equation}
    \text{D/G}_{\rm{X}} \equiv \frac{m_{\rm{Si}}N_{\rm{Si,model}}\left(1 - 10^{[\rm{Si/X}]}\right)}{m_{\rm{H}}N_{\rm{H,tot}}}\,,
\end{equation}
where X refers to a non-depleted element. We find that our dust-to-gas ratio relative to sulfur for the SC is $\log~{\rm D/G}_{\rm{S,\,SC}}~=~-3.72^{+0.39}_{-0.46}$ and $\log {\rm D/G}_{\rm{O,\,SC}} \gtrsim -4.25$. Both of these measurements are  well within  the upper bounds of the SC dust provided by \cite{2024ApJ...966...76M}.

It is significant that we are able to detect levels of dust to which FIR emission is not sufficiently sensitive. emission (e.g., \citealt{2016A&A...586A.121L}; \citealt{2012AJ....143...82W}. The dust fraction which we measure in the SC is significantly larger than we would expect for halo gas at similar distances from the Galactic plane. UV-based gas depletion can therefore provide crucial bounds to the dust fraction of HVCs, where FIR emission is not sufficiently sensitive.

We find a significant level of silicon gas-phase depletion ([Si/S]~=~\sisSLA) within the SC relative to the surrounding halo gas ([Si/S]~=~$-0.21$ at $z=-1.3\kpc$). This is the first detection of dust depletion within the SC and is among the highest levels of gas-phase depletion which have been found in an HVC. Our observations are in agreement with previous bounds on dust depletion provided for the SC. Moreover, UV-based gas depletion measurements can provide a higher level of sensitivity to the dust fractions of HVCs, given that it is an indirect way to measure dust. Future studies can use our measurement of dust to better understand the history of the SC, and other observational work followup can characterize the gas-phase depletion within a wider array of HVCs.

\section{Summary}\label{section:summary}

In this study, we used absorption-line spectroscopy along two new QSO sightlines and photoionization modeling to probe quantities of metallicity and dust depletions in the SC. We summarize our results below:

\begin{enumerate}

    \item \textit{Dust Depletion:} For Sightline~D, we find a $3\sigma$ level silicon gas-phase depletion of [Si/S]~=~\sisSLA\, and $\text{[Si/O]}_{3\sigma}~ \lesssim$~\sioSLA. This translates to a dust-to-gas ratio of $\log \rm{D/G} \gtrsim -4.25$, which is consistent with previous work and provides important bounds to dust content in this HVC.
    \item \textit{Metallicity Distribution:} For the absorber along Sightline~D, we find a metallicity of $[{\rm S/H}]=\,$\shSLA$\,\pm0.15$. We additionally find an IVC with super-solar metallicity along Sightline~E of $[{\rm S/H}]=\,$\shSLB$\,\pm0.15$. If this absorption is associated with the SC, then it is the highest measured metallicity for this HVC to date. However, at this Galactic longitude, there could be confusion with foreground gas in the SAMW, which limits our confidence in its association with the SC. 
    \item \textit{Mixing with Halo Gas:} The low \hi\, column density of $\log(N_{\rm{\hi}}/\cmsq) = 18.56\pm0.25$, low volume density of $\log(n_{\rm{H}}/\rm{cm}^{-3}) = -1.62^{+0.10}_{-0.26}$,  and high Si\textsc{~ii} Doppler parameter $b=28.2\pm2.4\,\kms$  along Sightline D imply that the gas in the SC's adjacent cloudlets are in an active stage of mixing and ionization. The dust content along this sightline is significantly higher than expected for halo gas.  If the gas is mixing with the ambient medium, the dust would be diluting, implying that the main body of the SC could have a higher dust content.
\end{enumerate}

 The metallicity we measure along Sightline~D is in agreement with previous UV-based measurements along the interface between the SC and the Galactic halo (see Table~\ref{table:weighted_Zs} and \citealt{2016ApJ...816L..11F}). Because UV absorption is so sensitive to diffuse gas (e.g., \citealt{2023ApJ...946L..48F}), we were able to determine a dust-to-gas ratio value using saturated O\textsc{~i} absorption and strong Si\textsc{~ii} absorption. Together, this HVC's metallicity and dust content implies at least some of this HVC's gas has a galactic origin.

\acknowledgments
Support for this program was provided by NASA through through the grant \textit{HST}-GO-15161 from the Space Telescope Science Institute, which is operated by the Association of Universities for Research in Astronomy, Incorporated, under NASA contract NAS\,5-26555. Additional support for Horton was provided by NSF grant 2334434. This paper includes \textit{HST} data from programs 13840 (PI Fox) and 15161 (PI Barger) that are available through the Mikulski Archive for Space Telescopes (MAST: \dataset[10.17909/fp16-9z51]{http://dx.doi.org/10.17909/fp16-9z51}) and \hi~21-cm data from GBT under program 23A-344 (PI Wakker). We also wish to acknowledge the atomic transition database from the National Institute of Standards and Technology (NIST; \citealt{2020Atoms...8...56R}), without which our work would not be possible. \\
\facilities{Hubble Space Telescope, Green Bank Telescope}
\software{\textsc{AstroPy} \citep{2022ApJ...935..167A}, \textsc{VoigtFit} \citep{krogager2018}, \textsc{MatPlotLib} \citep{Hunter:2007}, \textsc{NumPy} \citep{harris2020array}, and \textsc{Cloudy} v23.00
\citep{2023RMxAA..59..327C}.} 

\bibliographystyle{aasjournal} 
\bibliography{arxiv_version} 

\end{document}